\documentclass[10pt,twoside]{IEEEtran}

\usepackage{colortbl}
\usepackage{cite}
\usepackage{amsmath} 
\usepackage{amssymb} 
\usepackage{dsfont}
\usepackage[mathcal]{euscript}
\usepackage{empheq}
\usepackage{graphicx}
\usepackage{graphics}
\usepackage{bm}
\usepackage{psfrag}
\usepackage{mathrsfs}
\usepackage{bbm}
\usepackage{color}
\usepackage{cancel}
\usepackage{amsthm}
\input{epsf}

\newtheorem{definition}{Definition}
\newtheorem{assumption}{Assumption}

\newcommand{\E}{\mathbb{E}}

\newcommand{\beq}{\begin{equation}}
\newcommand{\eeq}{\end{equation}}
\newcommand{\beqa}{\begin{eqnarray}}
\newcommand{\eeqa}{\end{eqnarray}}
\newcommand{\dfz}{\triangleq}

\begin{document}

\title{Graph Learning Under Partial Observability}

\author{Vincenzo~Matta, Augusto~Santos, and Ali~H.~Sayed
\thanks{V.~Matta is with DIEM, University of Salerno,
via Giovanni Paolo II, I-84084, Fisciano (SA), Italy (e-mail: vmatta@unisa.it).

A. Santos was with the Adaptive System Laboratory, EPFL, CH-1015 Lausanne, Switzerland (e-mail: augusto.pt@gmail.com).

A.~H.~Sayed is with the \'Ecole Polytechnique F\'ed\'erale de Lausanne EPFL, School of Engineering, CH-1015 Lausanne, Switzerland (e-mail: ali.sayed@epfl.ch).
}
\thanks{
The work of A.~H.~Sayed was supported in part by grant 205121-184999 from the Swiss National Science Foundation.
}
}

\maketitle

\begin{abstract}
Many optimization, inference and learning tasks can be accomplished efficiently by means of {\em decentralized processing algorithms} 
where the network topology (i.e., the {\em graph}) plays a critical role in enabling the interactions among neighboring nodes. There is a large body of literature examining the effect of the graph structure on the performance of decentralized processing strategies. In this article, we examine the inverse problem and consider the reverse question: How much information does observing the behavior at the nodes of a graph convey about the underlying topology? 
For large-scale networks, the difficulty in addressing such inverse problems is compounded by the fact that usually only a {\em limited} fraction of the nodes can be probed, giving rise to a second important question: Despite the presence of {\em unobserved} nodes, can partial observations still be sufficient to discover the graph linking the probed nodes? 
The article surveys recent advances on this challenging learning problem and related questions. 
\end{abstract}


\section{Introduction}

\IEEEPARstart{T}{his} survey deals with complex systems whose evolution is dictated by interactions among a large number of elementary units (referred to as {\em network nodes}). The interactions give rise to some form of decentralized information processing that is characterized by two fundamental features: $i)$ the locality of information exchange between the individual units; and $ii)$ the capability to solve rather effectively a range of  demanding tasks (such as optimization, learning, and inference) that would otherwise be unattainable by stand-alone isolated nodes. 

There is a large body of literature that examines how the graph topology linking the nodes affects the performance of decentralized processing methods --- see, e.g.,~\cite{TsitsiklisBertsekasAthansTAC1986, NedicBertsekasSIAM2001, NedicOzdaglar2010, BoydFoundTrends, NedicJSTSP2013, KhanTAC2016, KhanTAC2018, RabbatRibeiroCoopGraphSP2018, BajwaTSIPN, SayedTuChenZhaoTowficSPmag2013,SayedProcIEEE2014,Sayed,ChenSayedTIT2015part1, ChenSayedTIT2015part2}. 
This article focuses on the reverse question, namely, what information the optimization solution conveys about the underlying topology. Specifically, assuming that we are able to observe the evolution of the signals at a subset of the nodes, we would like to examine what type of information can be extracted from these measurements in relation to the interconnections between the nodes.

\begin{figure} [t]
\begin{center}
\includegraphics[scale= 0.27]{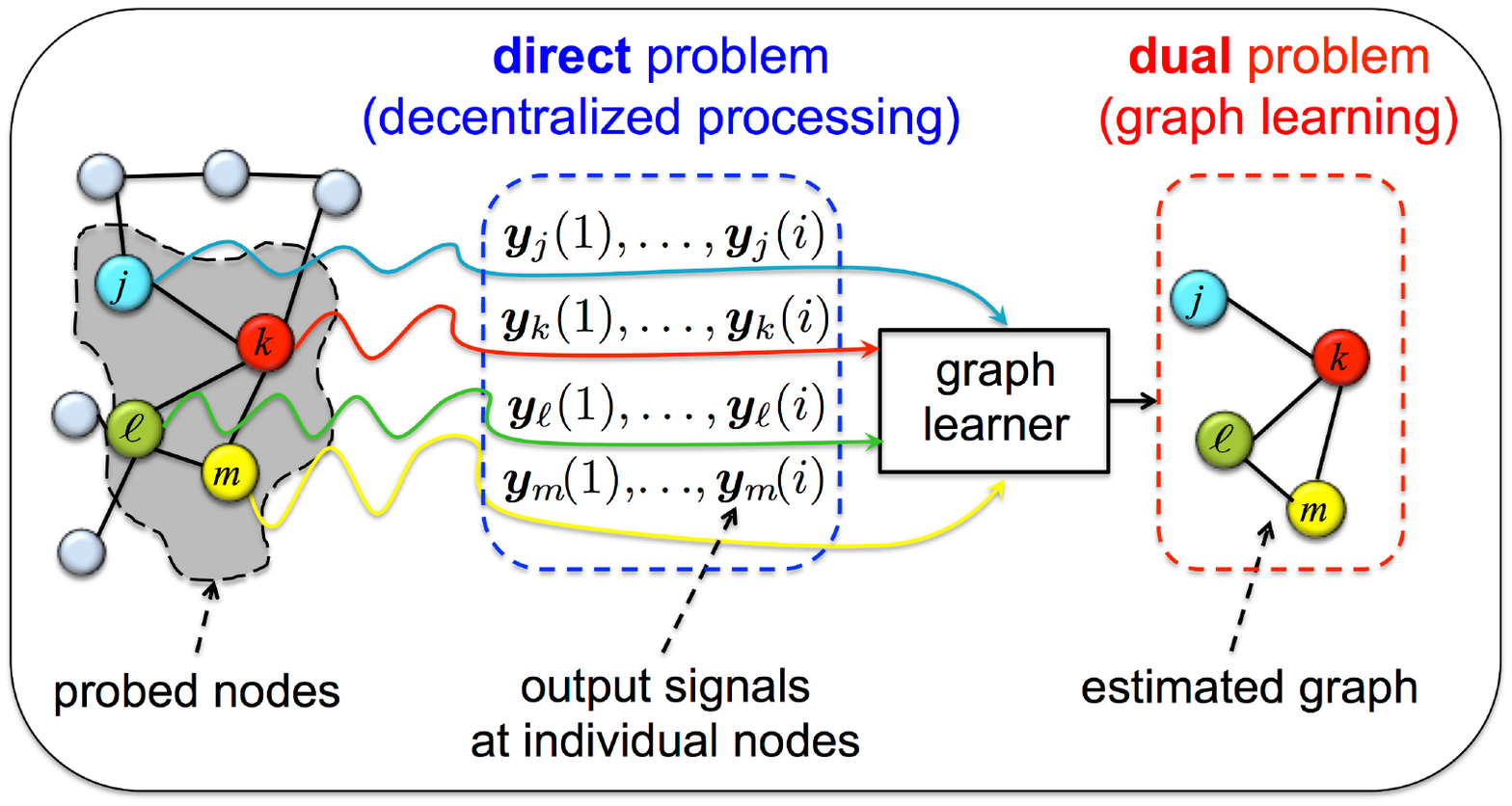}
\caption{Illustration of the graph learning problem considered in this article. 
A network performs a decentralized processing task (the {\em direct} learning problem). 
The network graph influences the way each node exchanges information with its neighbors. The online output of the decentralized processing at node $k$ and time $i$ is denoted by $\bm{y}_k(i)$. 
An inferential engine can probe the subset $\{j,k,\ell,m\}$ of the network, and collect the pertinent outputs. 
Based on these outputs, the goal of the {\em dual} learning problem is to estimate the subgraph of connections between nodes $j,k,\ell,m$.}
\label{fig:scheme1}
\end{center}
\end{figure}

Rather than focus on {\em what} the nodes learn through decentralized processing (which is the goal of the {\em direct} learning problem), we focus instead on a {\em dual} learning problem that deals with {\em how} the nodes learn (i.e., on discovering the hidden interconnections that drive the learning process).
A schematic illustration of this combined interplay is provided in Fig.~\ref{fig:scheme1}. In the direct problem, we start from a graph topology, run a decentralized processing algorithm, and analyze its performance (such as convergence rate and closeness to optimal solution) and the dependence of this performance on the graph. In the dual problem, we start from observing the signals generated by the nodes and focus instead on discovering the underlying graph that led to the observed signal evolution. 

The graph learning problem has many challenging aspects to it, as we explain below. Nevertheless, it is a problem of fundamental importance arising across a variety of application domains and disciplines\footnote{Since the considered problem arises across multiple disciplines, it is referred to in multiple ways including: graph learning, topology inference, network tomography, graph reconstruction, graph estimation. In the following, we will mostly use ``graph learning''.} because it can provide answers to many useful questions of interest. For instance, by observing the evolution of signals at a subset of the nodes, can one establish which nodes are sharing information with each other? Or how is privacy reflected in the nodes' signals? Also, by observing the behavior of some nodes, can one discover which nodes are having a magnified influence on the overall behavior of the network? 
Applications that can benefit from such answers are numerous. For example, discovering who is communicating with whom over the Internet~\cite{topoepidemics, Pedrosource, VenkitasubramaniamHeTongTIT2008, EmbeddingCapacityTIT2013}; tracing the information flow over a social network to capture the mechanism of opinion formation or to locate the source of fake news~\cite{KrimetalTSIPN2016,SLTL2019}; using graphs to characterize the evolution of urban traffic~\cite{DeriMouraTraffic}; learning the synchronized cognitive behavior of a school of fish evading predators~\cite{Couzin2009,Partridge1982}; investigating the relationship between structural and functional connectivity in the brain~\cite{BrainTomo}. 

In this article we focus on networks governed by {\em discrete-time linear dynamical systems} described by Eq.~(\ref{eq:mainmod}) further ahead.
This class of models has found application in many domains. For example, it is a classical model used in economics for time-series forecasting of financial data~\cite{Hamilton}; it has been applied in biostatistics and bioinformatics to estimate gene regulatory networks from gene expression data~\cite{generegul}; it arises automatically over networks deployed to solve distributed inference tasks, such as distributed detection problems~\cite{CattivelliSayedTSP2011,MattaSayedCoopGraphSP2018}.

There exist some useful survey articles related to the topic of graph learning~\cite{GiannakisProcIEEE,mateos,RabbatSPM2019}. 
However, most prior works assume that all nodes in a network are monitored. This is usually not the case. For example, in probing signals from the brain, only certain localities are examined. Also, in probing signal flows over a social network with millions of members, only a limited number of observations may be available. It is therefore critical to study how these limitations influence the problem of graph learning. As such, a core feature of this article is that we deal with {\em large} networks. 
Over these networks, due to different forms of physical limitations, it is not practical to assume that data can be collected from all nodes. This is seldom the case and our standing assumption in this presentation will be that observations are collected from only a subset of the nodes. We refer to this scenario as the {\em partial observability} regime. As a result, the graph learning task becomes more complicated than usual, since the observations collected at the monitored nodes are influenced (through information propagation) by the unobserved nodes. 
It is then natural to inquire whether this partial observability setting leads to an ill-posed graph learning problem or can still provide sufficient information to learn the underlying graph linking the observed nodes. 
This is a hard problem, which will not be feasible in general.

The main aim of this article is to survey some recent advances on graph learning under partial observability for networks governed by linear dynamical systems. In particular, we will find that, {\em despite the (massive, since the network is large) influence of the latent unobserved nodes}, the topology of the monitored subnetwork can be estimated well under proper conditions, and we will illustrate the meaning of these conditions. The roadmap we follow to pursue this goal is as follows.
In Sec.~\ref{sec:probform} we start by formulating the problem, then in Sec.~\ref{sec:literature} we illustrate the main issues in graph learning and how they are dealt with in the literature. 
We then focus on some recent theoretical advances in the field, which show how graph learning under partial observability can be feasible, in a setting that considers random graphs and certain properties of the combination matrix that the nodes employ in the evolution of the distributed network algorithms --- see Sec.~\ref{sec:ERgraphs}. 
Section~\ref{sec:illexam} is devoted to illustrating graph learning in operation: first we present a distributed detection example; then we use the experiments to highlight useful properties of the graph learning algorithms such as their complexity, performance and finite-size effects; and finally we show how an overall graph can be learned by sequentially reconstructing smaller portions thereof. Conclusions and perspectives follow in Sec.~\ref{sec:conclu}.

\vspace*{5pt}
{\bf Notation}. 
We use boldface letters to denote random variables, and normal font letters for their realizations. 
Matrices are denoted by capital letters, and vectors by small letters. 
This convention can be occasionally violated, for example, the total number of network nodes is denoted by $N$.
A random vector $\bm{x}$ that depends on a spatial (i.e., node) index $k$ and a time index $i$ is denoted by $\bm{x}_{k,i}$. 
A (scalar) random variable that depends on a spatial index $k$ and a time index $i$ is denoted by $\bm{x}_{k}(i)$. 
The symbol $\stackrel{\textnormal{p}}{\longrightarrow}$ denotes convergence in probability as $N\rightarrow\infty$. 
When we say that an event occurs ``w.h.p.'' we mean that it occurs ``with high probability'' as $N\rightarrow\infty$. 
Sets and events are denoted by upper-case calligraphic letters, whereas the corresponding normal font letters will denote set cardinalities. For example, the cardinality of $\mathcal{S}$ is $S$. The complement of $\mathcal{S}$ is denoted by $\mathcal{S}'$. 
For a $K\times K$ matrix $Z$, the submatrix spanning the rows of $Z$ indexed by set $\mathcal{S}\subseteq\{1,2,\ldots, K\}$ and the columns indexed by set $\mathcal{T}\subseteq\{1,2,\ldots,K\}$, is denoted by $Z_{\mathcal{S} \mathcal{T}}$, or alternatively by $[Z]_{\mathcal{S} \mathcal{T}}$. When $\mathcal{S}=\mathcal{T}$, the submatrix $Z_{\mathcal{S} \mathcal{T}}$ is abbreviated as $Z_{\mathcal{S}}$. 
The symbol $\log$ denotes the natural logarithm.

\section{Formulation of the Problem}
\label{sec:probform}
We are given a connected network of $N$ nodes, which implement a distributed diffusion algorithm in the form that we are going to illustrate. 
The output of node $k=1,2,\ldots,N$ at time $i\geq 0$ will be henceforth assumed to be a {\em random} variable denoted by $\bm{y}_k(i)$. 
For a given time instant, the outputs of all nodes are stacked into an $N\times 1$ column vector:
\beq
\bm{y}_i=[\bm{y}_1(i),\bm{y}_2(i),\ldots,\bm{y}_N(i)]^{\top}.
\eeq 
Likewise, a second random variable $\bm{x}_{k}(i)$ will represent the input data (or some function thereof), giving rise to the vector:
\beq
\bm{x}_i=[\bm{x}_1(i),\bm{x}_2(i),\ldots,\bm{x}_N(i)]^{\top}.
\eeq 
We assume that the input variables $\bm{x}_k(i)$ are independent and identically distributed (i.i.d.) both spatially (i.e., across node index $k$) and temporally (i.e., over time index $i$). 
We focus on the following diffusion model, a.k.a. first-order Vector AutoRegressive (VAR) model, which represents the diffusion learning process across the network:
\beq
\boxed{
\bm{y}_i=A \bm{y}_{i-1} + \bm{x}_i
}
\label{eq:mainmod}
\eeq
Expanding~(\ref{eq:mainmod}) on an entrywise basis we get:
\beq
\bm{y}_{k}(i)=\sum_{\ell=1}^N a_{k\ell} \bm{y}_{\ell}(i-1) + \bm{x}_{k}(i).
\label{eq:VARexpand}
\eeq
We see from~(\ref{eq:VARexpand}) that the structure of the {\em combination matrix} $A=[a_{k\ell}]$ is critical in determining how node $k$ incorporates information coming from node $\ell$. 
In particular, the skeleton of $A$ (i.e., the support graph given by the locations of the nonzero entries of $A$) encodes the possible paths that the information can follow through during the diffusion process~(\ref{eq:VARexpand}). 

In the graph learning problem under partial observability, only a limited subset $\mathcal{S}$ of nodes can be probed (i.e., their signals $\{\bm{y}_k(i)\}_{k\in\mathcal{S}}$ observed), and the main goal is to estimate the support graph $G_{\mathcal{S}}$ of the submatrix $A_{\mathcal{S}}$ (recall that this notation refers to restricting $A$ to the rows and columns defined by the indices in $\mathcal{S}$). 
The graph learning pipeline can be summarized in the following flow diagram: 
\beq
\boxed{
\begin{array}{cccccc}
\underbrace{
\bm{Y}=\{\bm{y}_{k}(1),\bm{y}_{k}(2),\ldots,\bm{y}_{k}(i)\}_{k\in\mathcal{S}}
}_{\textnormal{gather signals from $\mathcal{S}$}}
\\
\Downarrow
\\
\underbrace{\widehat{\bm{A}}_{\mathcal{S}}={\sf f}(\bm{Y})}_{\textnormal{estimate the combination submatrix in $\mathcal{S}$}}
\\
\Downarrow
\\
\underbrace{\widehat{\bm{G}}_{\mathcal{S}}={\sf h}(\widehat{\bm{A}}_{\mathcal{S}})}_{\textnormal{estimate the subgraph in $\mathcal{S}$}}
\end{array}
}
\label{eq:GLpipeline}
\eeq
In~(\ref{eq:GLpipeline}), the function ${\sf f}$ represents a mapping from the data to an estimated combination submatrix, while the function ${\sf h}$ is a suitable {\em thresholding} or {\em clustering} operator that classifies the entries of $\widehat{\bm{A}}_{\mathcal{S}}$ as connected/disconnected.

According to~(\ref{eq:GLpipeline}), one fundamental step is to devise a suitable function ${\sf f}$ to estimate the combination matrix. On first thought, it may appear natural to choose ${\sf f}$ as the covariance matrix, since one expects that the statistical correlation between the signals at two nodes provides an indication on whether they are connected or not. 
On closer reflection, however, one finds that this approach is problematic and more effective solutions are necessary. 
This is because over a connected network with cooperative nodes, pairwise correlation between two nodes is also affected by data streaming from other nodes through the successive local interactions: Nodes interact with their neighbors, which in turn interact with their neighbors, and so forth. 
As a result, if node $k$ is connected to $\ell$ through an intermediate node $m$, the outputs of $k$ and $\ell$ will be correlated even though there is no direct link between them. 
For this reason, it is not true in general that the {\em combination matrix entries can be faithfully estimated from the corresponding covariance matrix entries.}
This is true only for special networks that are called correlation networks, but many other possibilities exist. 
For example, in a Gaussian graphical model~\cite{Whittaker}: $i)$ the measurements at the network nodes obey a multivariate normal distribution with a certain covariance matrix; and $ii)$ the nonzero entries of the {\em inverse} of the covariance matrix (a.k.a. concentration matrix) correspond to the support graph of the network. 
But it should be remarked that even this result is not general enough, and that effective estimators for the graph must necessarily depend as well on the signal dynamics over the graph. 
The next section focuses on suitable choices for the model in~(\ref{eq:mainmod}).

\subsection{Estimating the Combination Matrix $A$ in Model~(\ref{eq:mainmod})}
For ease of presentation, in the forthcoming treatment we will assume, without loss of generality, that the random variables $\{\bm{x}_k(i)\}$ in~(\ref{eq:mainmod}) are zero mean and have unit variance. 
Multiplying both sides of~(\ref{eq:mainmod}) by $\bm{y}_{i-1}^{\top}$ and taking expectations, we obtain:
\beq
\underbrace{\E\left[\bm{y}_i \bm{y}_{i-1}^{\top}\right]}_{\stackrel{i\rightarrow\infty}{\longrightarrow} R_1}=
A \underbrace{\E\left[\bm{y}_{i-1} \bm{y}_{i-1}^{\top}\right]}_{\stackrel{i\rightarrow\infty}{\longrightarrow} R_0}
+
\underbrace{\E[\bm{x}_i \bm{y}_{i-1}^{\top}]}_{=0},
\label{eq:AR0R1}
\eeq
where the last term is zero because the sequence $\{\bm{x}_i\}$ is formed by independent and zero-mean random vectors, 
and where $R_0$ and $R_1$ are the limiting covariance and one-lag covariance matrices, respectively (these limits exist if $A$ is a stable matrix)~\cite{Sayed}. 
From~(\ref{eq:AR0R1}) we immediately observe that the matrix $A$ can be expressed as:
\beq
A=R_1 R_0^{-1},
\label{eq:Granger0}
\eeq
which is relevant for graph learning because covariance matrices can be estimated from samples, with increasing accuracy as the number of samples increases.
The solution in~(\ref{eq:Granger0}) can be interpreted as searching for the coefficients $\{a_{k\ell}\}$ that provide the best (in mean-square-error sense) linear prediction of $\bm{y}_{i}$ given the past sample $\bm{y}_{i-1}$ --- see, e.g.,~\cite{Wiener1956}. 
Estimator~(\ref{eq:Granger0}) is also known as Granger estimator or predictor, a terminology that arises in the context of Granger causality~\cite{Granger1969}.\footnote{In a nutshell, Granger causality refers to the relationships between time series. With reference to our example, assume that we regress $\bm{y}_{k}(i)$ on the past one-lag time series available in the network, $\bm{y}_{\ell}(i-1)$, for $\ell=1,2,\ldots,N$. 
As we have observed, the optimal predictor minimizing the regression error would not use the time series with $a_{k\ell}=0$ to predict $\bm{y}_{k}(i)$. Thus, one says that $k$ is ``Granger-caused'' by those $\ell$ for which $a_{k\ell}\neq 0$.}

However, in order to evaluate $R_0$ and $R_1$ we need to probe the {\em entire} network. 
Accordingly, estimator~(\ref{eq:Granger0}) is not useful under the partial observability regime adopted here, where only nodes belonging to subset $\mathcal{S}$ are probed. 
One approach to estimate the submatrix $A_{\mathcal{S}}$ could be by applying~(\ref{eq:Granger0}) to the covariance submatrices pertaining to $\mathcal{S}$:
\beq
\widehat{A}_{\mathcal{S}} = [R_1]_{\mathcal{S}} ([R_0]_{\mathcal{S}})^{-1}.
\eeq
This approach would correspond to determining the coefficients $\{a_{k\ell}\}$ (for $k,\ell\in\mathcal{S}$) that provide the minimum-mean-square-error linear prediction of the subvector containing the elements of $\{\bm{y}_k(i)\}$ for $k\in\mathcal{S}$, given the subvector of the past samples $\{\bm{y}_k(i-1)\}$ for $k\in\mathcal{S}$. 
Unfortunately, matrix analysis tells us that~\cite{Johnson-Horn}:
\beq
\boxed{
A_{\mathcal{S}}=
\left[R_1 R_0^{-1} \right]_{\mathcal{S}}
\neq [R_1]_{\mathcal{S}} ([R_0]_{\mathcal{S}})^{-1}
}
\label{eq:AsnotGranger}
\eeq
The middle term corresponds to extracting the ${\mathcal S}$ component from the product $R_1 R_0^{-1}$, whereas the last term corresponds to first extracting the ${\mathcal S}$ components from the individual covariances $R_1$ and $R_0$. The inequality sign is because the term $\left[R_1 R_0^{-1} \right]_{\mathcal{S}}$ takes into account the effect of the latent nodes {\em before} projection onto the set $\mathcal{S}$.
Therefore, a Granger predictor that ignores the latent variables is not necessarily satisfactory. 
In particular, the elementary result in~(\ref{eq:AsnotGranger}) provides an immediate hint on the fact that the graph learning problem is not necessarily feasible under partial observability.

\section{Literature Survey}
\label{sec:literature}

It is useful to illustrate three fundamental issues arising in the context of graph learning.

\subsection{Achievability, Hardness, and Sample Complexity}
I. {\em Achievability}.  
We say that graph learning is achievable when the graph of interest can be estimated well\footnote{We will quantify the qualification ``well'' in Sec.~\ref{sec:ident}, where we introduce formal notions of consistency to measure the accuracy of a graph estimate as the network size increases.} at least in the case of unlimited complexity. 
In this case, practical complexity constraints are ignored, for example, it is assumed that one can collect as many samples as desired and that the computational complexity associated with matrix inversion or search algorithms is not of concern.
To illustrate this concept, consider model~(\ref{eq:mainmod}) under full observability. From~(\ref{eq:Granger0}) we see that graph learning is achievable since there is a closed-form relationship that allows retrieving $A$ from $R_0$ and $R_1$, and since we assume that the covariance matrices can be estimated perfectly from the data for a large number of samples. 
In our {\em partial observability} setting, achievability is a critical and challenging issue, due to the assumption that we can collect data from only a limited portion of the network, whereas the number of unobserved nodes may scale to infinity.
Fortunately, it has been shown that, under certain conditions, graph learning with partial observations is achievable~\cite{tomo,tomo2,tomo3,JalaliSanghaviICML2012}, as we will discuss in Sec.~\ref{sec:ERgraphs}. 
However, even when achievability is established, there are at least two other aspects to consider related to {\em hardness} and {\em sample complexity}.

II. {\em Hardness or Computational Complexity}.
When examining hardness, we continue to disregard the complexity associated with the number of samples. 
That is, we continue to assume that an infinite collection of samples is available, such that no error arises from statistical fluctuations and the statistical quantities of interest are perfectly known.
The concept of hardness is then related to the {\em computational} complexity required to determine the support graph. 
For instance, with reference to the model in~(\ref{eq:mainmod}), with infinitely many samples we can assume that $R_0$ and $R_1$ are perfectly known. Hence, hardness refers to the computational complexity required to estimate the support graph from $R_0$ and $R_1$, which essentially amounts to inverting a large matrix.  
In some other graph learning problems hardness becomes a serious issue, since an NP search would be required to estimate the graph~\cite{Shimony, graphlearnhard1, complexity_markov, Bento2009, Breslerhardness}.

III. {\em Sample Complexity}.
This concept refers to the number of samples that are required to perform accurate graph learning. It also relates to how the number of necessary samples scales with the dimensionality of the problem (i.e., the network size). 
The issue of establishing how limited sample availability affects the learning performance is particularly relevant in the {\em high-dimensional} setting where the number of samples can be significantly smaller than the network size, as happens in the theoretical domain of high-dimensional graphical models~\cite{Whittaker}, or in application domains such as gene regulatory networks~\cite{generegul}. 

It is useful to illustrate the sample complexity issue in relation to problems where one estimates covariance matrices (e.g., under Gaussian graphical models or VAR models).  
Empirical covariance matrices are known to be rank deficient when the number of samples is smaller than or equal to the network size, which is clearly a problem when one needs to estimate the concentration matrix (inverse of the covariance matrix), or when one needs to compute a Granger estimator like the one in~(\ref{eq:Granger0}).
Even when the empirical covariance is not singular, the number of samples necessary to attain satisfactory performance can be large.
For example, as we will see later, the nonzero entries of the combination matrix usually become smaller as the network size increases. This means that for large networks, it becomes necessary to increase the accuracy of the empirical covariance matrices. When possible, one may resort to structural constraints (such as sparsity or smoothness) to regularize the estimation of the covariance matrices and keep sample complexity under control.  
One useful technique over sparse graphical models is the {\em graphical} LASSO method to estimate the concentration matrix~\cite{Tibshirani}. 

The majority of results that are available for sample complexity in the context of graphical models do not apply to graphs obeying dynamical systems like~(\ref{eq:mainmod}). This is because most of these results assume graphical models with i.i.d. observation samples rather than observations that arise from a dynamical model with memory~\cite{mateos}.
Some results on the sample complexity associated with model~(\ref{eq:mainmod}) appear in~\cite{BentoIbrahimiMontanari,LohWainwrightAOS2012, RaoKipnisJavidiEldarGoldsmithCDC2016,HanLuLiuJMLR} but they refer to the setting of full observability. Under partial observability, the issue is considered in~\cite{tomo3,JalaliSanghaviICML2012}.

\subsection{Graph Learning Under Full Observability}
Owing to the nature of model~(\ref{eq:mainmod}), we will mainly focus on {\em linear} system dynamics, but hasten to add that there exist works on graph learning over nonlinear dynamical systems as well~\cite{GiannakisProcIEEE,Napoletani, Noise_Jien_Ren, Koopman_Gon, Ching2017ReconstructingLI, PhysRevE.lai, ScienceRobustNetInference}.

One useful work on graph learning over linear systems is~\cite{MaterassiSalapakaTAC2012}, which considers a fairly general class of models (including {\em non-causal} systems and VAR models of any order). The main contribution of~\cite{MaterassiSalapakaTAC2012} is to devise an inferential strategy relying on Wiener filtering to retrieve the network graph. 
Such strategy is shown to guarantee exact reconstruction for the so-called {\em self-kin} networks. 
For more general network structures, the reconstruction of the smallest self-kin network embodying the true network is guaranteed.

In the context of graph signal processing~\cite{OrtegaSPmag, KovacevicSP2015, TsitsiveroBarbarossaDiLorenzo2016, PerraudinVandergheynstSP2017, ChepuriLeusTSIPN2017}, recent works focus on autoregressive diffusion models of arbitrary order~\cite{SantiagoTopo, pasdeloup, MeiMoura}. 
As a common feature of many of these works, the estimation algorithms leverage some prior knowledge about the graph structure, which is then translated into appropriate structural constraints. Typical constraints are in terms of sparsity of the connections, or smoothness (in the graph signal terminology) of the signals defined at the graph nodes. 
In~\cite{SantiagoTopo}, a two-step inferential process is proposed, where: $i)$ a graph shift operator~\cite{MouraDSP1,MouraDSP2,MarquesSegarraLeusRibeiroSP2017} is estimated through the nodes' signals that arise from the diffusion process; and $ii)$ given the {\em spectral templates} obtained from this estimation, the eigenvalues that would identify the graph are then estimated by adding proper structural constraints (e.g., sparsity) that could render the problem well-posed. 
In~\cite{pasdeloup}, the same concept of a two-step procedure is considered, with the main goal being to characterize the space of valid graphs, namely, graphs that can explain the signals measured at the network nodes. 
In~\cite{MeiMoura}, a model for {\em causal graph processes} is proposed, which exploits both inter-relations among nodes' signals and their intra-relations across time. Capitalizing on these relations, a viable algorithm for graph structure recovery is designed, which is shown to converge under reasonable technical assumptions.

There also exist works on graph learning over other types of dynamical systems. 
In~\cite{Kiyavash1}, a graphical model is proposed to represent networks of stochastic processes. Under suitable technical conditions, it is shown that such graphs are consistent with {\em directed information graphs}, which are based on a generalization of Granger causality. It is proved how directed information quantifies causality in a specific sense and efficient algorithms are devised to estimate the topology from the data. 
In~\cite{Kiyavash2}, a novel measure of causality is introduced, which is able to capture functional dependencies exhibited by certain (possibly nonlinear) network dynamical systems. 
These dependencies are then encoded in a functional dependency graph, which becomes a representation of possibly directed (i.e., causal) influences that are more sophisticated than the classical types of influences encoded in linear network dynamical systems.

Results for graph learning over continuous-time linear dynamical systems described by stochastic differential equations are provided in~\cite{BentoIbrahimiMontanari}. Conditions to achieve consistent graph learning are offered, along with a sample complexity analysis that relies on concentration bounds for the empirical covariance matrix. A least-squares algorithm with $\ell_1$-norm regularization is proposed. The analysis in~\cite{BentoIbrahimiMontanari} goes through a discretization of the model, which can be relevant also to the analysis of discrete-time diffusion models like the one in~(\ref{eq:mainmod}). 
For these latter models, achievability of consistent graph learning over {\em sparse} graphs is examined in~\cite{HanLuLiuJMLR }. 
An algorithm is designed, which tries to fit (\ref{eq:AR0R1}) with the most sparse matrix possible.
Some generalizations of this result to the case of missing observations are offered in~\cite{LohWainwrightAOS2012, RaoKipnisJavidiEldarGoldsmithCDC2016}, where samples from the entire network are gathered, but they can be intermittently available, or corrupted (these available observations are called ``partial observations'', but the meaning is different from the one adopted in this article, since in~\cite{LohWainwrightAOS2012, RaoKipnisJavidiEldarGoldsmithCDC2016} all nodes are probed, and the qualification ``partial'' refers to intermittence of observations at each node).

In summary, the aforementioned works (which we list with no pretense of exhaustiveness) address under various settings the problem of achievability and complexity of graph learning under the {\em full} observability regime. 
However, we must recall that in our setting we focus on the {\em partial} observability regime where a large portion of the network is not accessible. Most challenges in terms of feasibility of the graph learning problem will in fact stem from this complication.

\subsection{Graph Learning under Partial Observability}
In the presence of unobserved network nodes, there are results allowing proper graph learning when the topology is of some pre-assigned type (polytrees)~\cite{MaterassiSalapakaCDC2012, KiyavashPolytrees}.
For fairly arbitrary graph structures, some results about the possibility of correct graph retrieval are provided in~\cite{Geigeretal15, MaterassiSalapakaCDC2015}.
One limitation of these results resides in the fact that the sufficient conditions for graph learning depend on some ``microscopic'' details about the model (e.g., about the local structure of the topology or the pertinent statistical model). 
For this reason, over large-scale networks (which are the focus of this article) a different approach is necessary.  

One approach suited to large networks is an asymptotic analysis carried out as the network size $N$ scales to infinity. 
In order to cope with the large network size in a way that enables a tractable analysis, it is useful to model the network graph as a {\em random} graph. An asymptotic analysis can then become feasible, letting emerge the {\em thermodynamic} properties of the graph, with the conditions for graph learning being summarized in some {\em macroscopic} (i.e., average) indicators, such as the probability that two nodes of the random graph are connected.

Similar forms of asymptotic analysis were recently performed for {\em high-dimensional} graphical models with latent variables. 
In~\cite{AnandkumarTanHuangWillskyJMLR2012}, the focus is on Gaussian graphical models, and consistent graph learning is proved (along with a viable algorithmic solution) under an appropriate {\em local separation criterion}. 
In~\cite{AnandkumarValluvanAOS2013} results of consistent learning are instead provided for {\em locally tree-like} graphs. 
Graph learning under the so-termed ``{\em sparsity+low-rank}'' condition is examined in~\cite{ChandrasekaranParriloWillskyAOS2012}. Under this condition (where the observed subnetwork is sparse and the unobserved subnetwork is low-rank in an appropriate sense), it is proved that the graph and the amount of latent variables can be jointly estimated. 
Moreover, in~\cite{AnandkumarTanHuangWillskyJMLR2012,AnandkumarValluvanAOS2013,ChandrasekaranParriloWillskyAOS2012}, a detailed analysis of sample complexity is provided, which is especially relevant since these works focus on the {\em high-dimensional} setting where the number of samples can be smaller than the network size.
In~\cite{BreslerBoltzmann}, a graphical model consisting of a ferromagnetic restricted Boltzmann machine with bounded degree is considered. It is shown that such class of graphical models can be effectively learned through the usage of a novel influence-maximization metric.

However, classical graphical models (such as the ones used in the aforementioned references) do not assume that there are {\em signals} evolving over time at the network nodes. In contrast, classical graphical models assume a still picture of the network, where the data measured at the individual nodes are modeled as random variables characterized by a certain joint distribution. 
The inter-node statistical dependencies are encoded in the joint distribution through an underlying graph.
Under this framework, estimation of the graph from the data defined at the nodes is performed assuming that the inferential engine has access to i.i.d. samples of these data, and there is no model of the evolution of the data across time. 

For this reason, the results obtained in the aforementioned references on graph learning in the presence of latent variables do not apply to the dynamical system considered in~(\ref{eq:mainmod}). 
Results relevant to the latter system are provided in~\cite{JalaliSanghaviICML2012}, starting from the ``{\em sparsity+low-rank}'' approach proposed in~\cite{ChandrasekaranParriloWillskyAOS2012}. In~\cite{JalaliSanghaviICML2012} it is assumed that the probed subgraph is sparse, and that a certain matrix associated with the unobserved nodes is low-rank, which in particular means that the number of unobserved nodes must be smaller than the number of probed ones.  
In order to fit~(\ref{eq:mainmod}), a regularized least-squares algorithm is proposed, where $\ell_1$-norm regularization is used to control sparsity, and nuclear-norm regularization to control the rank of the matrix associated with the latent network part.

Exploiting the properties of Erd\H{o}s-R\'enyi (ER) random graphs and the regularity of the combination matrices used in typical distributed processing settings, some recent advances provide examples of achievable graph learning under partial observability when the graph of probed nodes is not necessarily sparse, and the number of latent nodes can be arbitrarily large~\cite{tomo, tomo2, tomo3, tomo_icassp, tomo_dsw, tomo_isit, MattaSantosSayedAsilomar2018, MattaSantosSayedISIT2019}. 
The forthcoming section summarizes these advances in some detail.

\section{Achievable Graph Learning}
\label{sec:ERgraphs}

As explained in the previous section, for large networks it is necessary to perform some asymptotic analysis to obtain useful analytical results, and to establish the fundamental thermodynamic properties that emerge with high probability over the network. 
One typical way to tackle this problem is to {\em randomize} the network structure, i.e., to work with {\em random graphs}. 
One useful class of random graphs is the celebrated model proposed by Erd\H{o}s and R\'enyi~\cite{erd,BollobasRandom}, which is an undirected graph where the probability that nodes $k$ and $\ell$ are connected is a Bernoulli random variable characterized by a certain connection probability $p$, and where all edges are drawn independently and with the same connection probability.

An important graph descriptor is the {\em degree} of a node. 
The degree of node $k$ is defined as the number of neighbors of node $k$ (including $k$ itself), and will be denoted by $\bm{d}_k$. 
Owing to the Bernoulli model, the average degree ${\sf D}_{\textnormal{av}}$ of every node in an Erd\H{o}s-R\'enyi graph is equal to $1+(N-1) p$.

\subsection{Graph Evolution Regimes}
Let us examine the evolution of the random graph when $N$ grows. When the connection probability is a constant $p>0$, the number of neighbors increases linearly with $N$ (in the following, the notation $\sim$ means ``scales as'', when $N\rightarrow\infty$):
\beq
\boxed{
{\sf D}_{\textnormal{av}}\sim N p\qquad\textnormal{[dense regime]}
}
\label{eq:Davpconst}
\eeq  
It is not difficult to figure out that, since in this case any node has a number of neighbors growing as $N$, the graph exhibits a {\em dense} connection structure, and for sufficiently large $N$, is likely to be a connected graph, i.e., a graph where there always exists an undirected path connecting any pair of nodes. 
However, a fundamental result from random graph theory states that, in order to ensure a graph is connected with high probability as $N$ grows, the minimal growth of the average degree is~\cite{erd,BollobasRandom}:
\beq
\boxed{
{\sf D}_{\textnormal{av}}=\log N + \mathcal{O}(\log N)\qquad \textnormal{[log-sparse regime]}
}
\label{eq:DavpN}
\eeq
where by $\mathcal{O}(\log N)$ we denote a sequence that diverges\footnote{The Big-O notation $f(N)=O(g(N))$ usually means that $|f(N)|$ is upper bounded by $c |g(N)|$ for some constant $c$ and sufficiently large $N$. Our notation $\mathcal{O}(f(N))$ adds the requirement that $f(N)\rightarrow+\infty$ as $N\rightarrow\infty$.} to $+\infty$ at most logarithmically and, hence, the connection probability $p\sim{\sf D}_{\textnormal{av}}/N$ vanishes. 
The logarithmic growth corresponds in fact to a {\em phase transition}, since it represents the minimal growth that ensures a connected graph. 

There is yet a third (sparse) connected regime, which is intermediate between the log-sparse and the dense regimes introduced so far. This intermediate regime occurs when the average degree grows faster than logarithmically (while the connection probability still vanishes), formally when: 
\beq
\boxed{
{\sf D}_{\textnormal{av}}=\omega_N \log N\qquad \textnormal{[intermediate-sparse regime]} 
}
\label{eq:strongconc}
\eeq
where $\omega_N\rightarrow\infty$ in an arbitrary fashion, but sufficiently slow so as to ensure that the connection probability vanishes.

\begin{figure} [t]
\begin{center}
\[
\begin{array}{cc}
\includegraphics[scale= 0.48]{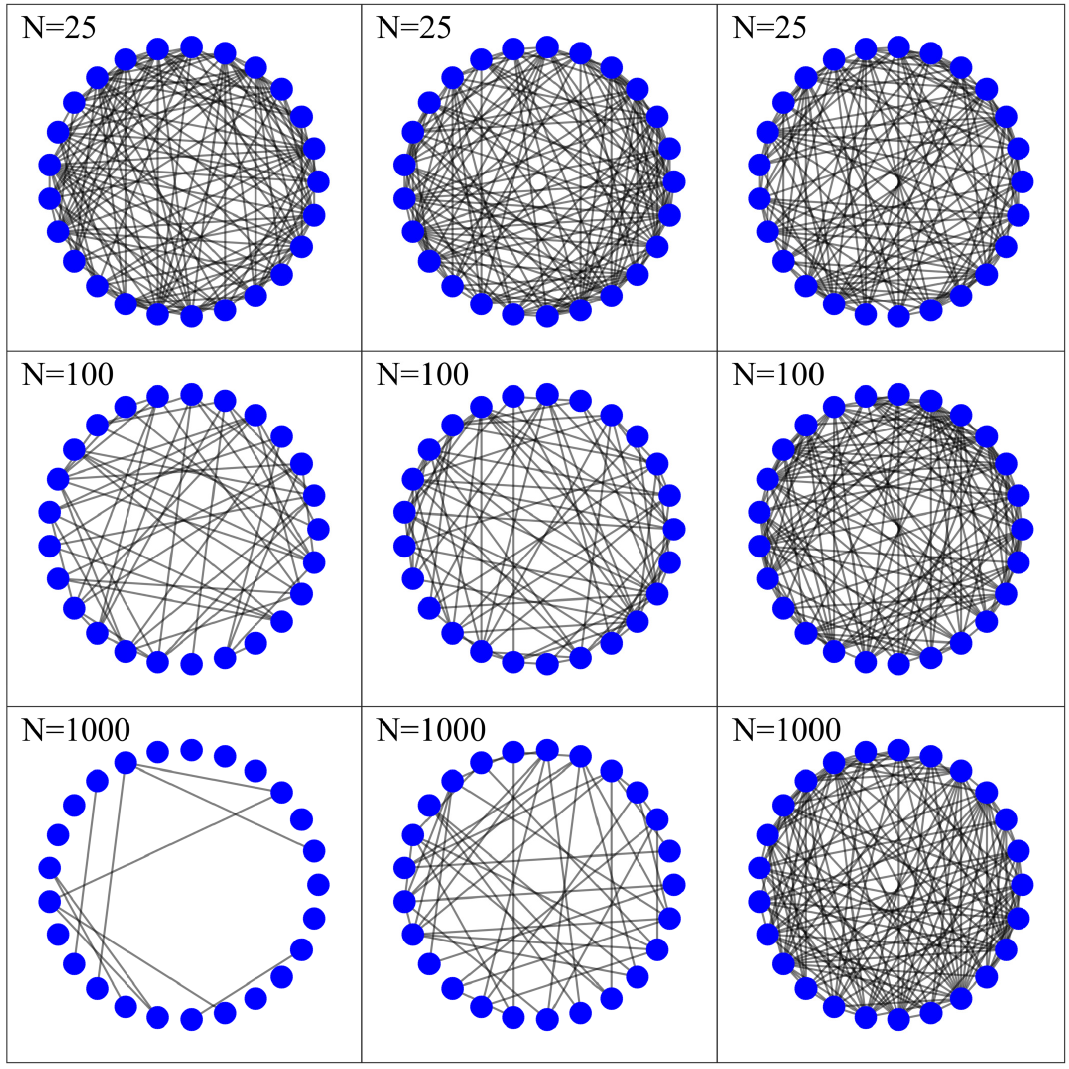}\\
\includegraphics[scale= 0.235]{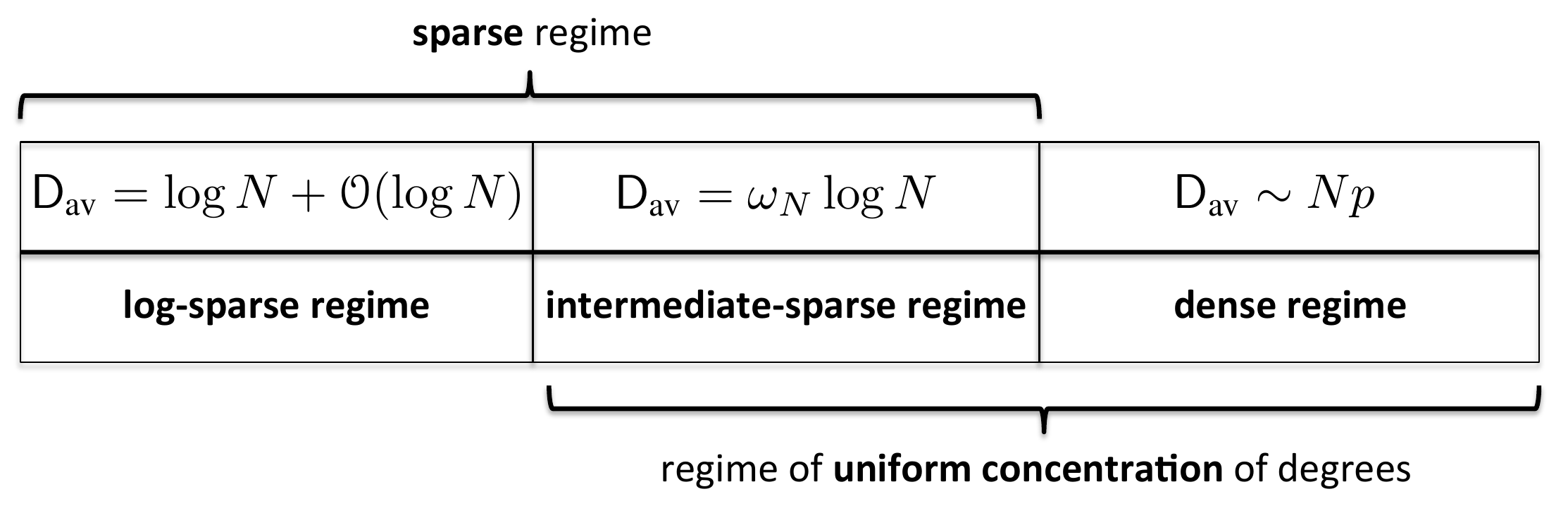}
\end{array}
\]
\caption{
Taxonomy of the {\em connected} regimes for the ER model. The overall {\em sparse} regime is given by the union of the log-sparse and intermediate-sparse regimes. In comparison, the union of the intermediate-sparse and dense regimes gives rise to the {\em uniform concentration of degrees}, which will be seen to play an important role in the graph learning problem addressed in this article. 
Each column of the plot grid corresponds to a different regime, from left to right: sparse, intermediate-sparse, and dense, respectively. Moving across rows, we consider networks of increasing (from top to bottom) total number of nodes $N$. For clarity of visualization, in all panels we display only the subgraph of the first $25$ nodes of the network.
}
\label{fig:Venn}
\end{center}
\end{figure}

There is one fundamental property that holds under the intermediate-sparse and dense regimes, but not under the log-sparse regime, and is the following {\em statistical concentration} property:\footnote{We note that the term ``concentration'' does {\em not} refer to the number of node connections, but, according to a standard terminology adopted in statistics, refers to statistical quantities that collapse to some deterministic value as $N\rightarrow\infty$~\cite{ConcIneqBook}.} 
\beq
\boxed{
\frac{\bm{d}_{\min}}{{\sf D}_{\textnormal{av}}}\stackrel{\textnormal{p}}{\longrightarrow} 1,~
\frac{\bm{d}_{\max}}{{\sf D}_{\textnormal{av}}}\stackrel{\textnormal{p}}{\longrightarrow} 1,~ \textnormal{[uniform degree concentration]}
}
\label{eq:dmaxmin}
\eeq
where $\bm{d}_{\min}$ and $\bm{d}_{\max}$ denote the minimal and maximal degree over all nodes, respectively. 
This means that, under~(\ref{eq:dmaxmin}), the minimal and maximal degree concentrate around the expected degree. 

The overall taxonomy comprising the different elements of sparsity, density, and degree concentration, is reported in Fig.~\ref{fig:Venn}, along with an example of evolution, as $N$ grows, of the ER graphs corresponding to the different regimes. 
For each regime, we consider an ER graph of increasing size ($N=25, 100, 1000$), and for each value of $N$ we display (for clarity of visualization) the behavior of the first $25$ nodes of the network. 
For all regimes we start with a connection probability equal to $1/2$. 
Accordingly, the top panels have similar shape. 
Then, as $N$ increases, the connection probability obeys the scaling law relative to the particular regime. 
In the leftmost panels (sparse regime), we see that the displayed subgraph becomes progressively more sparse.\footnote{We remark that the overall graph, which is too large to be displayed, remains connected even if the shown subgraph becomes progressively disconnected. In fact, on the overall graph with $N$ nodes, we can leverage the increasing number of nodes to find a path between any two nodes (with high probability) provided that the connection probability scales appropriately.}
In the middle panels (intermediate-sparse regime), sparsity increases, but some more structure is preserved. 
Finally, in the rightmost panels (dense regime), the subgraph has an invariant behavior.

We see that the union of the log-sparse and intermediate-sparse regimes identifies the sparse (as opposed to the dense) regime. Likewise, the union of the intermediate-sparse and dense regimes identifies the regime of uniform degree concentration.

\subsection{Partial Observability Settings}
The main challenge of the graph learning problem considered in this article is related to the partial observability setting, where only a subset $\mathcal{S}$ of the network can be probed. 
In order to deal with the asymptotic regime, it is necessary to define how the cardinality $S$ scales with the overall network size $N$. In particular, we introduce the {\em asymptotic fraction of probed nodes} $\xi$: 
\beq
\boxed{
\frac{S}{N}\stackrel{N\rightarrow\infty}{\longrightarrow} \xi
}
\label{eq:cardscaling}
\eeq 
The extreme case where the cardinality of probed nodes is fixed when $N\rightarrow\infty$ corresponds to a {\em low-observability}  regime ($\xi=0$) where the set of unobserved nodes becomes dominant and infinitely larger than the subset of probed nodes. 
However, when the size of $\mathcal{S}$ is fixed and finite, it is not useful to model the connections within $\mathcal{S}$ through an ER model because, in the sparse regime, every edge in $\mathcal{S}$ would trivially disappear as $N$ gets large! 

In order to deal with the graph learning problem under the low-observability regime in a meaningful way, the following {\em partial} ER model was introduced in~\cite{tomo2}: $i)$ the subgraph of interest, $\mathcal{S}$, is {\em deterministic and arbitrary}; $ii)$ while the latent nodes act as a {\em noisy} disturbance, with the connections outside $\mathcal{S}$, and also between $\mathcal{S}$ and $\mathcal{S}'$ (the set of latent nodes), drawn according to an ER model. 

\begin{figure} [t]
\begin{center}
\includegraphics[scale= 0.2]{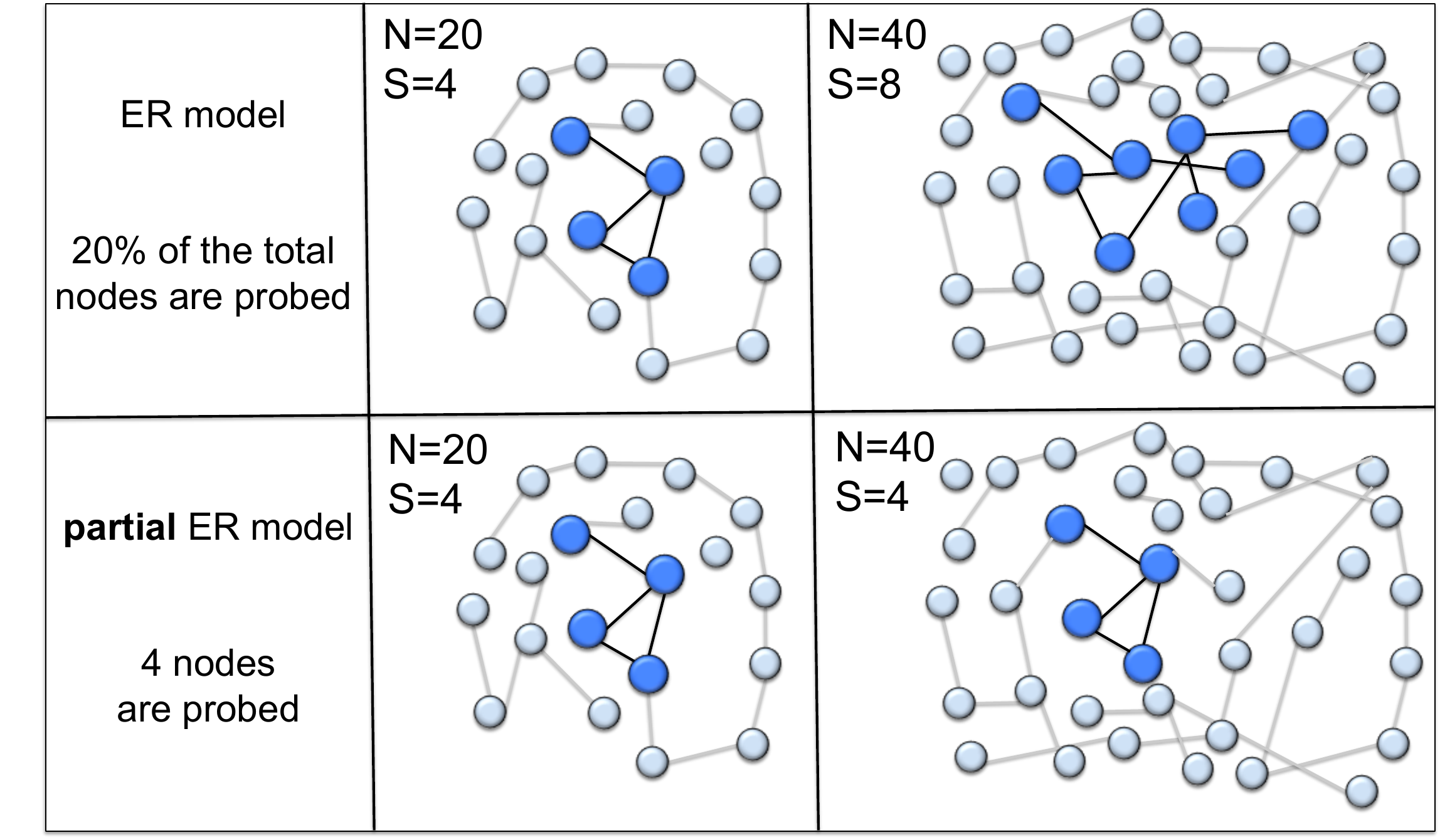}
\caption{
Partial observability settings considered in this article. The probed nodes forming the subgraph of interest are highlighted in blue.
The evolution for the plain ER regime is illustrated in the top panels. Here, as the network size grows (from $N=20$ to $N=40$), the number of probed nodes grows as well, with the fraction of probed nodes $\xi=S/N=0.2$ staying constant and with the subgraph of probed nodes varying.
The {\em partial} ER regime is illustrated in the bottom panels. Here, the number of probed nodes stays constant ($S=4$) as the network size grows and the structure of the probed subgraph is deterministically fixed. 
}
\label{fig:partialsettings}
\end{center}
\end{figure}

The distinction between the {\em plain} and {\em partial} ER models is illustrated in Fig.~\ref{fig:partialsettings}. 
In the top panels, a plain ER model with $\xi=0.2$ is considered. 
We see that the subset $\mathcal{S}$ of probed nodes (displayed in blue) increases from $S=4$ to $S=8$ when $N$ increases from $20$ to $40$. Moreover, the subgraph associated with $\mathcal{S}$ (as well as the overall graph) changes randomly its shape according to an ER model. 
In comparison, the {\em partial} ER model is displayed in the bottom panels. 
In this case, the subset $\mathcal{S}$ of probed nodes has {\em fixed} cardinality and structure. 
The edges (displayed in gray) between nodes belonging to the unobserved set $\mathcal{S}'$, as well as between $\mathcal{S}'$ and $\mathcal{S}$, are randomly drawn according to an ER model.

\subsection{Combination Matrices}

In the presence of {\em partial} observations, the graph learning problem can be ill-posed. 
In fact, while under full observability Eq.~(\ref{eq:Granger0}) guarantees that our inverse problem can be solved, under partial observability Eq.~(\ref{eq:AsnotGranger}) highlights that invertibility is lost due to the error introduced by unobserved nodes, and in general there are no guarantees that this error can allow accurate graph estimation. 
It makes sense to investigate whether it {\em can} for certain classes of combination matrices. 
In the following treatment, the matrix $A$ will be assumed to be symmetric and a scaled (stable) version of a doubly-stochastic matrix, namely,
\beq
\boxed{
A=A^{\top},~~a_{k\ell}\geq 0,~~\sum_{\ell=1}^N a_{k\ell}=\rho,~~0<\rho<1
}
\label{eq:combmat0}
\eeq
This structure is motivated by the typical implementation of combination matrices employed in distributed optimization and learning strategies, for example in {\em consensus}~\cite{DeGroot,XiaoBoydSCL2004}, {\em gossip} algorithms~\cite{BoydGhoshPrabhakarShahTIT2006, DimakisKarMouraRabbatScaglioneProcIEEE2010}, or {\em diffusion} algorithms~\cite{SayedTuChenZhaoTowficSPmag2013,SayedProcIEEE2014,Sayed,ChenSayedTIT2015part1, ChenSayedTIT2015part2}. 
In these distributed implementations, if node $k$ is connected to node $\ell$, it scales the output received from $\ell$ through some nonnegative weight $a_{k\ell}$. In order to perform a distributed averaging, the weight sums are usually kept constant, as in~(\ref{eq:combmat0}). We will examine an example of these distributed implementations in the detection application considered in Sec.~\ref{subsec:DAD}.

One useful qualification of the combination matrices in~(\ref{eq:combmat0}) that is relevant to graph learning is in terms of the {\em variability} of their nonzero entries. We introduce two pertinent classes for these matrices.
\begin{assumption}[Class $\mathcal{V}_1$]
\label{assum:regdiffmat}
The nonzero entries of the combination matrix, scaled by the average degree ${\sf D}_{\textnormal{av}}$, do not vanish, namely, given that $k$ and $\ell$ are connected, a certain $\theta>0$ exists such that, with high probability for large $N$:
\beq
\boxed{
{\sf D}_{\textnormal{av}}\,\bm{a}_{k\ell}>\theta
}
\label{eq:class1}
\eeq~\hfill$\square$
\end{assumption}
Condition~(\ref{eq:class1}) is motivated by the following observation. 
For typical choices of combination matrices, each node $k$ distributes the weight mass $\rho$ across its neighbors. Thus, we will have typically, over connected pairs $(k,\ell)$:
\beq
\bm{a}_{k\ell}\propto \frac{1}{{\sf D}_{\textnormal{av}}},
\eeq
which explains why the quantity ${\sf D}_{\textnormal{av}} \,\bm{a}_{k\ell}$ does not vanish, and why condition~(\ref{eq:class1}) is meaningful.
\begin{assumption}[Class $\mathcal{V}_2$]
\label{assum:regdiffmat2}
We assume that, for connected pairs $k$ and $\ell$: 
\beq
\boxed{
\frac{\kappa}{\bm{d}_{\max}}\leq
\bm{a}_{k\ell}\leq \frac{\kappa}{\bm{d}_{\min}}
}
\label{eq:fundamentalassum}
\eeq
for some $0<\kappa\leq \rho$.~\hfill$\square$
\end{assumption}
We see from~(\ref{eq:fundamentalassum}) that, when an edge exists connecting $k$ and $\ell$, the variation of the (nonzero) matrix entries is defined in terms of the (reciprocal of the) maximal and minimal graph degrees. Also this condition can be motivated by the observation that nodes tend to distribute the weights across their neighbors in some homogeneous way. It is possible to show that, under the connectivity regimes for the ER model considered here, the left inequality in~(\ref{eq:fundamentalassum}) implies~(\ref{eq:class1}), namely, we can conclude that~\cite{tomo}:
\beq
\mathcal{V}_2\subset\mathcal{V}_1.
\eeq 
That is, the conditions for a matrix to be in class $\mathcal{V}_2$ are more stringent than the conditions required to be in class $\mathcal{V}_1$.

As a matter of fact, the most popular combination matrices used in distributed processing belong to class $\mathcal{V}_2$ and, hence, to $\mathcal{V}_1$.
Two notable instances are the Laplacian and Metropolis combination rules, which can be defined as follows~\cite{Sayed}. 
For $k\neq\ell$, with $k$ and $\ell$ connected:
\beq
\begin{array}{lll}
&\bm{a}_{k\ell}=\displaystyle{\frac{\alpha\rho}{\bm{d}_{\max}}},~~~~~~~~&\textnormal{[Laplacian rule]}
\\
\\
&\bm{a}_{k\ell}=\displaystyle{\frac{\rho}{\max\left\{\bm{d}_k,\bm{d}_{\ell}\right\}}},&\textnormal{[Metropolis rule]}
\end{array}
\label{eq:LapMetroMat}
\eeq
For both rules, the self-weights are determined by the third condition in~(\ref{eq:combmat0}), yielding $\bm{a}_{kk}=\rho - \sum_{\ell\neq k} \bm{a}_{k\ell}$. For the Laplacian rule, the parameter $\alpha$ satisfies $0<\alpha\leq 1$.

\subsection{Consistent Graph Learning}
\label{sec:ident}
In the following, the term ``consistency'' refers to the possibility of learning the graph correctly as $N\rightarrow\infty$. 
We will see that different notions of consistency are possible. We start from the weakest one. 

We denote by $\widehat{\bm{A}}_{\mathcal{S}}$ a certain estimate for the combination (sub)matrix corresponding to subset $\mathcal{S}$. 
We explain in the next section several ways by which such an estimate can be computed. 
We remark that the consistency results presented next in Sec.~\ref{sec:summa} will hold for (plain or partial) ER graphs and symmetric combination matrices. 
Nevertheless, it is useful to formulate the general theory to handle more general types of graphs (also {\em directed}) and asymmetric combination matrices. 
For this reason, when we refer to node pairs we will actually mean {\em ordered} pairs, with $(k,\ell)$ being distinct from $(\ell,k)$, because a {\em directed} link could exist from $\ell$ to $k$ and not vice versa.

We first introduce a general thresholding rule to classify connected/disconnected pairs. We will declare that the ordered $(k,\ell)$ pair is connected (i.e., that the $(k,\ell)$ entry of the {\em true} combination matrix is nonzero) if the corresponding {\em estimated} matrix entry, $\widehat{\bm{a}}_{k\ell}$, exceeds some threshold $\tau$. Accordingly, let us introduce the following error quantities:
\beqa
\bm{\mathcal{E}}_0(\tau)&\dfz&\frac{\textnormal{no. of entries where $\bm{a}_{k\ell}=0$ and $\widehat{\bm{a}}_{k\ell}>\tau$}}{\textnormal{no. of entries where $\bm{a}_{k\ell}=0$}},
\nonumber\\
\bm{\mathcal{E}}_1(\tau)&\dfz&\frac{\textnormal{no. of entries where $\bm{a}_{k\ell}>0$ and $\widehat{\bm{a}}_{k\ell}\leq\tau$}}{\textnormal{no. of entries where $\bm{a}_{k\ell}>0$}},\nonumber\\
\label{eq:E01}
\eeqa
where we assume $(k,\ell)\in\mathcal{S}$ with $k\neq \ell$. 
More informally, Eqs.~(\ref{eq:E01}) can be rephrased as:
\beqa
\bm{\mathcal{E}}_0(\tau)&\dfz&\frac{\textnormal{no. of mistakenly classified disconnected pairs}}{\textnormal{no. of disconnected pairs}},
\nonumber\\
\bm{\mathcal{E}}_1(\tau)&\dfz&\frac{\textnormal{no. of mistakenly classified connected pairs}}{\textnormal{no. of connected pairs}}.\nonumber\\
\eeqa
\begin{definition}[Weak Consistency]
Let $\widehat{\bm{A}}_{\mathcal{S}}$ be a matrix estimator. We say that $\widehat{\bm{A}}_{\mathcal{S}}$ achieves weak consistency if there exists a threshold $\tau$ (possibly depending on $N$) such that:
\beq
\boxed{
\bm{\mathcal{E}}_0(\tau)+\bm{\mathcal{E}}_1(\tau)\stackrel{\textnormal{p}}{\longrightarrow} 0
}
\label{eq:weak}
\eeq~\hfill$\square$
\label{def:weak}
\end{definition}
The notion of consistency in~(\ref{eq:weak}) ensures that the {\em average} fraction of mistakenly classified edges goes to zero. 
When the cardinality $S$ of probed nodes is fixed (as happens in the low-observability regime with partial ER model), an average number of mistakes that goes to zero implies that the subgraph of $\mathcal{S}$ is perfectly recovered. 
In contrast, when the cardinality $S$ grows with $N$, ensuring a small {\em average} fraction of mistakes can be unsatisfactory, which motivates the qualification ``weak''. Let us clarify this issue through a simple example. 
Consider a reconstruction that is perfect, except for $100$ edges that are estimated by the learning algorithm but that are actually not present in the true graph. 
The average number of mistakes, $100/S$, goes to zero as the subnetwork size $S$ goes to infinity, but due to the $100$ spurious edges, we will never end up with perfect reconstruction. 
The presence of (even a small number of) spurious edges can be penalizing especially under the sparse regime, where the number of true edges is small, and a reconstructed network where the number of spurious edges is comparable with the number of true edges might be unsatisfactory.  

From these observations, we argue that stronger notions of consistency are desirable. To this aim, we now introduce the useful concepts of {\em margins} and {\em identifiability gap}~\cite{tomo3}.

\begin{figure} [t]
\begin{center}
\includegraphics[scale= 0.3]{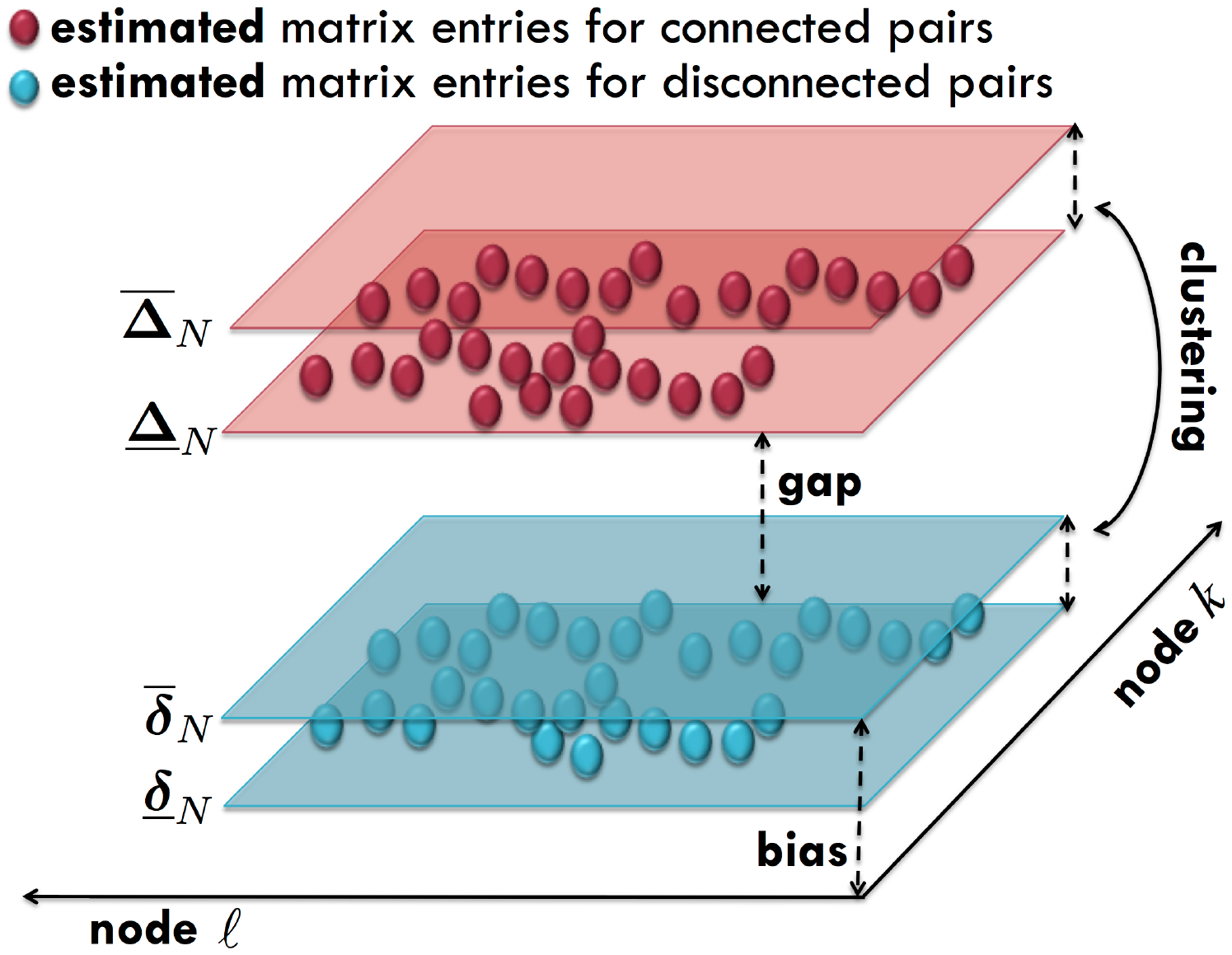}
\caption{Illustration of concepts useful for graph learning consistency. 
The estimated matrix entries corresponding to disconnected (resp., connected) pairs are sandwiched (clustering effect) between the margins $\underline{\delta}_N$ and $\overline{\delta}_N$ (resp., $\underline{\Delta}_N$ and $\overline{\Delta}_N$).
The separation between $\overline{\delta}_N$ and the origin determines the emergence of a bias. 
Likewise, the separation between $\underline{\Delta}_N$ and $\overline{\delta}_N$ determines the emergence of an identifiability gap. 
Technically, the definitions of $\eta$ and $\Gamma$ in~(\ref{eq:igap}) and~(\ref{eq:igapstrong}) require a scaling sequence $s_N$, which is not considered in the figure in order to convey the main idea without added complexity.
}
\label{fig:identifiability}
\end{center}
\end{figure}

\begin{definition}[Margins]
For a given matrix estimator $\widehat{\bm{A}}_{\mathcal{S}}$, we introduce the lower and upper margins corresponding to the disconnected pairs:
\beq
\underline{\bm{\delta}}_N
\dfz
\displaystyle{\min_{\substack{k,\ell\in\mathcal{S}: \bm{a}_{k\ell}=0 \\ k\neq \ell}}\widehat{\bm{a}}_{k\ell}},\quad
\overline{\bm{\delta}}_N
\dfz
\displaystyle{\max_{\substack{k,\ell\in\mathcal{S}: \bm{a}_{k\ell}=0 \\ k\neq \ell}}\widehat{\bm{a}}_{k\ell}},
\label{eq:upperlowerdisc}
\eeq
and the lower and upper margins corresponding to the connected pairs:
\beq
\underline{\bm{\Delta}}_N
\dfz
\displaystyle{\min_{\substack{k,\ell\in\mathcal{S}: \bm{a}_{k\ell}>0 \\ k\neq \ell}}\widehat{\bm{a}}_{k\ell}},\quad
\overline{\bm{\Delta}}_N
\dfz
\displaystyle{\max_{\substack{k,\ell\in\mathcal{S}: \bm{a}_{k\ell}>0 \\ k\neq \ell}}\widehat{\bm{a}}_{k\ell}}.
\label{eq:upperlowerconn}
\eeq~\hfill$\square$
\end{definition}
The physical meaning of the margins is to identify upper and lower bounds on the entries corresponding to node pairs of a given type (connected/disconnected). 
For example, the lower and upper margins for the disconnected pairs identify a region (see Fig.~\ref{fig:identifiability}) where we can find {\em all} the entries of the {\em estimated} matrix corresponding to disconnected pairs. 
A similar interpretation holds for the connected pairs. 
Now, one would expect that a good estimator exhibits the desirable property that $\widehat{\bm{a}}_{k\ell}$ goes to zero if nodes $k$ and $\ell$ are not connected. While it is legitimate to aspire for this property, a more careful analysis reveals that correct classification can still occur even if, over disconnected pairs $(k,\ell)$, the entries $\widehat{\bm{a}}_{k\ell}$  go to some nonzero value (i.e, if they have a bias). 
The important property to enable correct classification is that the region of disconnected pairs stays clear and separated from the region of connected pairs, which means that some {\em gap} must exist between the {\em upper} margin over disconnected pairs and the {\em lower} margin over connected pairs. This observation leads naturally to the definitions of {\em bias} and {\em gap}, and to the associated 
concept of {\em strong} consistency.
\begin{definition}[Strong Consistency]
Let $\widehat{\bm{A}}_{\mathcal{S}}$ be a matrix estimator. If there exist a positive sequence $s_N$, a real value $\eta$, and a positive value $\Gamma$, such that, for an arbitrarily small $\epsilon>0$:
\beq
\boxed{
\begin{array}{llll}
s_N \, \overline{\bm{\delta}}_N ~< \eta+\epsilon~~~~~~~~~\textnormal{w.h.p.}\\
s_N \,\underline{\bm{\Delta}}_N > \eta+ \Gamma - \epsilon ~~~~\textnormal{w.h.p.} 
\end{array}
}
\label{eq:igap}
\eeq
we say that $\widehat{\bm{A}}_{\mathcal{S}}$ achieves strong consistency, with a bias at most equal to $\eta$, an identifiability gap at least equal to $\Gamma$, and with a scaling sequence $s_N$.\footnote{We see that the definition of consistency includes a scaling sequence $s_N$. This scaling, which might look rather technical at first glance, admits a straightforward interpretation. For example, if we assume some homogeneity in the way the weights are distributed across the neighbors, the combination matrix entries over the connected pairs scale roughly as $1/{\sf D}_{\textnormal{av}}$ and, hence, they vanish as $N\rightarrow\infty$. Accordingly, it is necessary to scale them by $s_N={\sf D}_{\textnormal{av}}$ to get a stable asymptotic behavior.
}~\hfill$\square$ 
\label{def:strong}
\end{definition}

We remark that the latter concept of consistency is strong because it entails the possibility of recovering asymptotically {\em without errors} the true graph in $\mathcal{S}$. In fact, the separation between the region of connected and disconnected pairs implied by~(\ref{eq:igap}) suggests that proper classification can be performed by comparing the estimated matrix entries against some threshold comprised between $\eta/s_N$ and $(\eta+\Gamma)/s_N$ (disregarding the small $\epsilon$ for sufficiently large $N$).
It is nevertheless evident from~(\ref{eq:igap}) that, in order to evaluate the classification threshold, certain system parameters should be known beforehand. 
First of all, one should know the bias and the gap, and these quantities depend on several system parameters such as parameters of the combination matrix or the connection probability~\cite{tomo3}. Moreover, one should know the scaling sequence $s_N$. For example, if $s_N={\sf D}_{\textnormal{av}}$, one should be able to predict the average number of neighbors to set a proper threshold.
For these reasons, in practical applications, it will be more useful to have a data-driven mechanism (such as a clustering procedure) that would allow us to set the classification threshold automatically from the observed data. 
We will use the qualification ``universal'' to denote these data-driven techniques.
Accordingly, we can strengthen once more the notion of consistency to embody the universality requirement.
\begin{definition}[Universal strong consistency]
Let $\widehat{\bm{A}}_{\mathcal{S}}$ be a matrix estimator. If there exist a positive sequence $s_N$, a real value $\eta$, and a positive value $\Gamma$, such that:
\beq
\boxed{
\begin{array}{llll}
s_N \,\underline{\bm{\delta}}_N \stackrel{\textnormal{p}}{\longrightarrow} \eta,\qquad 
s_N \,\underline{\bm{\Delta}}_N \stackrel{\textnormal{p}}{\longrightarrow} \eta + \Gamma
\\
s_N \,\overline{\bm{\delta}}_N \stackrel{\textnormal{p}}{\longrightarrow} \eta,\qquad
s_N \,\overline{\bm{\Delta}}_N \stackrel{\textnormal{p}}{\longrightarrow} \eta+ \Gamma
\end{array}
}
\label{eq:igapstrong}
\eeq
we say that $\widehat{\bm{A}}_{\mathcal{S}}$ achieves universal strong consistency, with a bias $\eta$, an identifiability gap $\Gamma$, and with a scaling sequence $s_N$.~\hfill$\square$
\label{def:univ}
\end{definition}
We see from~(\ref{eq:igapstrong}) that the notion of {\em universal} strong consistency adds to the notion of strong consistency an inherent {\em clustering} ability. This is because the (scaled) margins corresponding to disconnected pairs, $s_N\,\underline{\bm{\delta}}_N$ and $s_N\,\overline{\bm{\delta}}_N$, converge to one and the same value, $\eta$, whereas the (scaled) margins corresponding to connected pairs, $s_N\,\underline{\bm{\Delta}}_N$ and $s_N\,\overline{\bm{\Delta}}_N$, converge to one and the same value, $\eta+\Gamma$. 
In light of this behavior, the estimated entries corresponding to disconnected pairs are squeezed to the bias $\eta$, and the estimated entries corresponding to connected pairs are squeezed to the higher value $\eta+\Gamma$, giving rise to two well-separated clusters that allow (asymptotically) faithful classification by means of a universal clustering algorithm --- see Fig.~\ref{fig:identifiability}. 

It is useful to compare~(\ref{eq:igapstrong}) against~(\ref{eq:igap}). 
We see that~(\ref{eq:igap}) does not require that the margins converge, but requires that the upper margin over disconnected pairs is confined below some value, and the lower margin over connected pairs is confined above some (higher) value. Unfortunately, the mere fact that the regions of connected and disconnected pairs are separated might not be sufficient to set the classification threshold from the data. In order to see why, consider a situation where the (scaled) entries below $\eta$ are separated in two clusters, and the (scaled) entries above $\eta+\Gamma$ are arbitrarily disposed. Then, in the absence of any prior information, an automated threshold procedure is likely to get confused since it cannot determine whether the two clusters below $\eta$ correspond to the same class or not. This unpleasant situation cannot occur if~(\ref{eq:igapstrong}) is verified.

\subsection{Relevant Matrix Estimators}
\label{sec:ME}
A general matrix estimator $\widehat{\bm{A}}_{\mathcal{S}}$ can always be written as:
\beq
\widehat{\bm{A}}_{\mathcal{S}}=\bm{A}_{\mathcal{S}}+\bm{E},
\label{eq:generalerror}
\eeq
where $\bm{E}$ is an error matrix. 
We see from the decomposition in~(\ref{eq:generalerror}) that there are two main ingredients to establish consistency for the graph learning problem. One is the asymptotic behavior of the {\em true} matrix $\bm{A}_{\mathcal{S}}$ (how do its entries scale when $N$ goes to infinity?).
Assume that there is a scaling sequence $s_N$ ensuring that the true entries over the connected pairs converge somewhere. 
Then, the asymptotic behavior of the error matrix $\bm{E}$ becomes critical. 
For example, if the error (scaled by $s_N$) converges to zero, then we can hope to recover the true graph, but other interesting situations can occur. In fact, according to what we illustrated in Sec.~\ref{sec:ident}, a nonzero error bias does not impair graph learning provided that a suitable gap between connected and disconnected pairs arises in the respective estimated matrix entries.

We now introduce three matrix estimators that have been recently applied to graph learning under partial observability~\cite{tomo,tomo2,tomo3}. 
Preliminarily, it is useful to observe that the steady-state self and one-lag covariance matrices in~(\ref{eq:AR0R1}) can be evaluated in closed form as follows~\cite{Sayed}:
\beq
\bm{R}_0=( I - \bm{A}^2 )^{-1},\qquad
\bm{R}_1=\bm{A}\bm{R}_0=\bm{A}( I - \bm{A}^2 )^{-1},
\label{eq:R1}
\eeq
where $I$ is the $N\times N$ identity matrix, and where we remark that the bold notation for $\bm{R}_0$ and $\bm{R}_1$ is due to the randomness of the matrix $\bm{A}$, which inherits the randomness of the underlying ER graph.

The Granger estimator, as discussed in the introduction, is obtained by replacing~(\ref{eq:Granger0}) with its counterpart over the monitored subset $\mathcal{S}$, i.e., by accounting only for the probed nodes while neglecting the effect of the latent nodes in $\mathcal{S}'$:
\beqa
\widehat{\bm{A}}^{\textnormal{(Gra)}}_{\mathcal{S}}&=&[\bm{R}_1]_{\mathcal{S}} ([\bm{R}_0]_{\mathcal{S}})^{-1}\nonumber\\
&=&
\bm{A}_{\mathcal{S}}+
\underbrace{
\bm{A}_{\mathcal{S}\mathcal{S}'} 
\left(I_{\mathcal{S}'} - [\bm{A}^2]_{\mathcal{S}'}\right)^{-1} 
[\bm{A}^2]_{\mathcal{S}'\mathcal{S}}}_{\textnormal{error term}}.
\label{eq:Gradesiredpluserror}
\eeqa
In~(\ref{eq:Gradesiredpluserror}), $I_{\mathcal{S}'}$ is the submatrix of the $N\times N$ identity matrix $I$, relative to subset $\mathcal{S}'$, and the representation of the error term comes from classical results on block matrix inversion --- see~\cite{Johnson-Horn,tomo}.

Due to the one-lag regression structure of~(\ref{eq:mainmod}), another possibility is to consider $[\bm{R}_1]_{\mathcal{S}}$ as an estimator for the combination matrix. 
In relation to the graph learning goal, one useful property is that, using~(\ref{eq:R1}), the covariance submatrix $[\bm{R}_1]_{\mathcal{S}}$ can be written as the matrix of interest, $\bm{A}_{\mathcal{S}}$, plus some higher-order powers of $\bm{A}$, namely,
\beq
\widehat{\bm{A}}^{\textnormal{(1-lag)}}_{\mathcal{S}}=[\bm{R}_1]_{\mathcal{S}}
=
\bm{A}_{\mathcal{S}} + \underbrace{[\bm{A}^3]_{\mathcal{S}} +[\bm{A}^5]_{\mathcal{S}} +\ldots}_{\textnormal{error term}}
\label{eq:onelagestdef}
\eeq

The third estimator is based on the (scaled) difference between consecutive time samples, which is sometimes referred to as the {\em residual}: $\bm{r}_i=(\bm{y}_i - \bm{y}_{i-1})/\sqrt{2}$.
Observing that $\E[\bm{r}_i \bm{r}_i^{\top}]=\bm{R}_0 - \bm{R}_1=(I_N + \bm{A})^{-1}$, we can introduce the matrix estimator:
\beq
\widehat{\bm{A}}^{\textnormal{(res)}}_{\mathcal{S}}=[\bm{R}_1]_{\mathcal{S}} - [\bm{R}_0]_{\mathcal{S}}+I_{\mathcal{S}}
=
\bm{A}_{\mathcal{S}} \underbrace{- [\bm{A}^2]_{\mathcal{S}} + [\bm{A}^3]_{\mathcal{S}} + \ldots}_{\textnormal{error term}}
\label{eq:newResidualSeries0}
\eeq

The asymptotic characterization of the error terms in~(\ref{eq:Gradesiredpluserror}),~(\ref{eq:onelagestdef}) and~(\ref{eq:newResidualSeries0})  was performed in~\cite{tomo,tomo2,tomo3}, yielding the achievability results summarized in the next section. In particular, the behavior of the three error matrices depends on the asymptotic behavior of the combination matrix powers (this structure is not evident in~(\ref{eq:Gradesiredpluserror}), and is shown in~\cite{tomo3}).

\begin{table*}[h]
  \begin{center}
    \begin{tabular}{|c|c|c|c|c|c|} 
      \hline
      \rowcolor[rgb]{0.86,0.86,0.86}
      &&&&&
      \\
      \rowcolor[rgb]{0.86,0.86,0.86}
      {\bf Result and reference}& {\bf Probed nodes} & {\bf Graph regime} & {\bf Matrix variability} & \bf{Consistency} & {\bf Estimator}\\
      \rowcolor[rgb]{0.86,0.86,0.86}
      &&&&&
      \\
      \hline
      &&&&&
      \\
      $T_1$ --- Ref.~\cite{tomo}&$\xi>0$ & sparse & class $\mathcal{V}_1$ & weak & Granger\\
      &&&&&
      \\
      \hline
      &&&&&
      \\
      $T_2$ --- Ref.~\cite{tomo2}&$\xi=0$ & sparse with $\displaystyle{\frac{\left(\log{\sf D}_{\textnormal{av}}\right)^2}{\log N}\rightarrow 0}$ & 
      class $\mathcal{V}_1$ & strong & Granger\\
      &&&&&
      \\
      \hline
      &&&&&
      \\
      $T_3$ --- Ref.~\cite{tomo3}&$\xi\geq 0$ & uniform concentration & class $\mathcal{V}_2$ & universal & Granger, one-lag, residual\\
      &&&&&
      \\
      \hline
    \end{tabular}
    \vspace*{5pt}
    \caption{Summary of the achievability results from~\cite{tomo,tomo2,tomo3}. 
    The condition on ${\sf D}_{\textnormal{av}}$ in the third column poses a slight limitation on the growth of the average degree, which implies that the result in~\cite{tomo2} covers the log-sparse and intermediate-sparse regimes, while not spanning entirely the latter.
    }
    \label{tab:summary}
  \end{center}
\end{table*}

\subsection{Summary of Results}
\label{sec:summa}
We now summarize some recent results for the problem of graph learning under partial observability~\cite{tomo,tomo2,tomo3}. 
The bottom line of the ensemble of these results is that graph learning under partial observability {\em is possible}. Our objective is to present the results in some unified way. 
Accordingly, we find it appropriate to avoid ``highly'' technical details and focus instead on the main insights. 
For each result, we direct the reader to the references where the technical details can be found.

The results in~\cite{tomo,tomo2,tomo3} differ in some aspects that can be summarized by the following five features. 
\begin{itemize}
\item {\em Fraction of probed nodes} $\xi$. This feature refers to the regime of observability. 
By writing $\xi=0$ we implicitly mean that we are focusing on the low-observability regime with partial ER model.
\item
{\em Graph regime}. This feature refers to the taxonomy in Fig.~\ref{fig:Venn}.
\item
{\em Matrix variability}. This feature refers to the matrix classes in Assumptions~\ref{assum:regdiffmat} and~\ref{assum:regdiffmat2}.
\item
{\em Consistency}. This feature refers to the notions of weak, strong, and universal consistency, reported in definitions~\ref{def:weak},~\ref{def:strong}, and~\ref{def:univ}, respectively.
\item
{\em Estimator}. This feature refers to the three types of estimators, namely, the Granger, the one-lag, and the residual estimators.
\end{itemize}
The main theorems available in~\cite{tomo,tomo2,tomo3} are compactly illustrated in Table~\ref{tab:summary}. We now briefly compare the three results $T_1$, $T_2$, $T_3$, as designated in the leftmost column of Table~\ref{tab:summary}.

Result $T_1$ from~\cite{tomo} shows that weakly-consistent graph learning with partial observations is possible under the sparse regime (either log-sparse or intermediate-sparse), for the case where the number of probed nodes $S$ grows with $N$ ($\xi>0$).
In order to overcome the limitations of weak-consistency --- see discussion following~(\ref{eq:weak}) ---  a refined analysis was conducted in~\cite{tomo} to examine the convergence rate of the errors in~(\ref{eq:weak}) and to show that the edges introduced in error by the estimation algorithm are asymptotically fewer than the true edges. However, the convergence analysis relies on some approximation and does not allow to conclude that the subgraph of interest is perfectly reconstructed as $N\rightarrow\infty$.

Result $T_2$ is from~\cite{tomo2}; the approach in this work differs from~\cite{tomo} and allows extending the results in two directions.
First, the challenging regime of {\em low observability} ($\xi=0$) is addressed, where the latent part becomes dominant, i.e., infinitely larger than the monitored part. Second, result $T_2$ is able to establish {\em exact reconstruction} since strong consistency is proved.

Results $T_1$ and $T_2$ pertain to the sparse regime. Result $T_3$ goes further and examines graph learning under the uniform concentration regime~\cite{tomo3,MattaSantosSayedAsilomar2018,MattaSantosSayedISIT2019}. Recall that the regime of uniform concentration is neither simply sparse nor dense, since it is defined as the union of the {\em intermediate-sparse} regime and the {\em dense} regime. 
Result $T_3$ exploits the asymptotic properties arising from the uniform degree concentration~(\ref{eq:dmaxmin}), coupled with the structure of combination matrices in class $\mathcal{V}_2$, to characterize the asymptotic behavior of the errors in~(\ref{eq:Gradesiredpluserror}),~(\ref{eq:onelagestdef}) and~(\ref{eq:newResidualSeries0}). 
We note that: 
\\
a) $T_3$ includes the {\em dense} case, and under this regime it provides guarantees of {\em universal} strong consistency. 
\\
b) Also in the intermediate-sparse regime $T_3$ provides guarantees of {\em universal} strong consistency, whereas $T_1$ and $T_2$ do not. However, $T_3$ holds for a more restricted class of matrices (class $\mathcal{V}_2$).
\\
c) $T_3$ cannot handle the log-sparse regime, which is instead handled by $T_1$ and $T_2$.
\\
d) $T_3$ shows consistency for two additional matrix estimators (which can be relevant in practice since they can deliver performance superior to the Granger estimator).

One relevant conclusion from $T_3$ is that, contrary to some widespread belief, sparsity is not necessarily the enabler of consistent graph learning. One fundamental element is seen to be the {\em uniform concentration of the graph degrees}, which coupled with the regular combination matrices in class $\mathcal{V}_2$ and the randomness of the ER model, gives rise to {\em universally strongly consistent} graph learning under partial observability. 
Nevertheless, sparsity has an impact on sample complexity, since it can be leveraged to make the estimation algorithms more efficient by introducing proper regularization constraints.

\section{Illustrative Examples}
\label{sec:illexam}

In all the forthcoming examples, we consider the graph learning procedure described in~(\ref{eq:GLpipeline}). In particular: $i)$ first, the combination matrix corresponding to the probed nodes is estimated using one of the three estimators presented in Sec.~\ref{sec:ME}; $ii)$ then, a (variant of the) $k$-means clustering algorithm is applied to classify the entries of the estimated matrix~\cite{tomo3}.  

\subsection{Distributed Detection}
\label{subsec:DAD}
One relevant application of distributed inference over networks is distributed detection, which can be formulated as follows~\cite{CattivelliSayedTSP2011,MattaSayedCoopGraphSP2018}. 
We are given a collection of streaming data $\{\bm{z}_{k}(i)\}$, where $k$ and $i$ are node and time indices, respectively. 
The data are both spatially and temporally i.i.d. according to two possible mutually exclusive hypotheses: the null hypothesis $H_0$ and the alternative hypothesis $H_1$, which correspond respectively to probability functions $\pi_0$ and $\pi_1$ (density functions for continuous variables, mass functions for discrete variables).  
We want to solve the detection problem in a distributed manner. To this end, we proceed as described in~\cite{MattaSayedCoopGraphSP2018}, and 
focus in particular on a diffusion implementation known as Combine-Then-Adapt (CTA), which is well-suited for learning from streaming data.
The CTA algorithm evolves by iterating the following two steps for every time $i$. 
First, during the {\em combination} step, every node $k$ computes an intermediate value $\bm{\psi}_{k}(i-1)$ as a weighted linear combination of the states $\{\bm{y}_{\ell}(i-1)\}$ arriving from its neighbors at previous time $i-1$:
\beq
\boxed{
\bm{\psi}_{k}(i-1)=\displaystyle{\sum_{\ell=1}^N c_{k\ell} \bm{y}_{\ell}(i-1)}~~\textnormal{[combine]}
}
\label{eq:CTACombine}
\eeq
In order to guarantee proper averaging, it is often assumed that the matrix $C=[c_{k \ell}]$ is doubly stochastic, meaning that it is nonnegative and that the entries on each of its rows and on each of its columns add up to $1$.

Second, during the {\em adaptation} step, each node uses its {\em locally-available current} data, $\bm{z}_{k}(i)$, to update the intermediate state from $\bm{\psi}_{k}(i-1)$ to the new state $\bm{y}_{k}(i)$. In particular, in detection applications, the update is performed based on the comparison between the old value $\bm{\psi}_{k}(i-1)$ and the local log-likelihood ratios, $\bm{\lambda}_k(i)=\log\frac{\pi_1(\bm{z}_{k}(i))}{\pi_0(\bm{z}_{k}(i))}$, of the fresh observations~\cite{MattaSayedCoopGraphSP2018}:
\beq
\boxed{
\bm{y}_{k}(i)=\bm{\psi}_{k}(i-1) - \mu\,\Big[
\bm{\psi}_{k}(i-1) - \bm{\lambda}_k(i)
\Big]~~ \textnormal{[adapt]}
}
\label{eq:CTAdapt}
\eeq
The scalar $\mu\in(0,1)$ appearing in~(\ref{eq:CTAdapt}) is commonly referred to as the {\em step-size}~\cite{Sayed}. 
The adaptation step has the purpose of taking into account the effect of the streaming data, allowing the system to track fast possible nonstationarities and drifts arising in these data. For example, if the underlying hypothesis changes over time, it is desirable that the distributed detectors recognize these changes. 

By introducing the $N\times 1$ vectors:
\beq
\bm{x}_i=
\mu\,\left[
\bm{\lambda}_{1}(i),\bm{\lambda}_{2}(i),\ldots,\bm{\lambda}_{N}(i)
\right]^{\top},
\eeq
and applying~(\ref{eq:CTACombine}) and (\ref{eq:CTAdapt}) in cascade, we get the useful matrix representation:
\beq
\bm{y}_i=(1-\mu) C \bm{y}_{i-1} + \bm{x}_i,
\label{eq:VARex1}
\eeq
which corresponds to~(\ref{eq:mainmod}) with the choice $A=(1-\mu) C$.

It is possible to show that, provided sufficient time for learning is granted, the individual states $\bm{y}_{k}(i)$ of each node will fluctuate (for sufficiently small $\mu$) around the expected value of the log-likelihood ratio. This expected value depends on the particular hypothesis in force, and is equal to (we suppress indices $k$ and $i$ due to identical distribution and denote by $\E_{\pi}$ expectation computed under distribution $\pi$):
\beqa
\textnormal{under } H_0&:&\E_{\pi_0}\left[\log\frac{\pi_1(\bm{z})}{\pi_0(\bm{z})}\right]=
-\mathcal{D}_{01},\nonumber\\
\textnormal{under } H_1&:&\E_{\pi_1}\left[\log\frac{\pi_1(\bm{z})}{\pi_0(\bm{z})}\right]=+\mathcal{D}_{10},
\eeqa
where $\mathcal{D}_{h j}$ denotes the Kullback-Leibler (KL) divergence between $\pi_h$ and $\pi_j$, for $h,j=0,1$~\cite{CT}. 
Accordingly, the output of each node will fluctuate around a negative or positive value depending on whether the true hypothesis is $H_0$ or $H_1$. Effective discrimination between the hypotheses can be attained through a decision rule that compares the output of the CTA algorithm against a threshold $\gamma\in (-\mathcal{D}_{01}, \mathcal{D}_{10})$. 
A fundamental tradeoff arises~\cite{Sayed,MattaSayedCoopGraphSP2018}: the smaller $\mu$ is, the smaller the size of the oscillations around the expected log-likelihood, which corresponds to a higher detection precision, but at the price of slower adaptation. In particular, the error probabilities scale {\em exponentially} fast with $1/\mu$ --- see~\cite{MattaBracaMaranoSayedTIT2016,MattaBracaMaranoSayedTSIPN2016} for a detailed asymptotic analysis.

\begin{figure} [t]
{\footnotesize{\bf ~~~~~~~~~~~~~~~distributed detection~~~~~~~~~~~~~graph learning
}}
\begin{center}
\includegraphics[scale= 0.4]{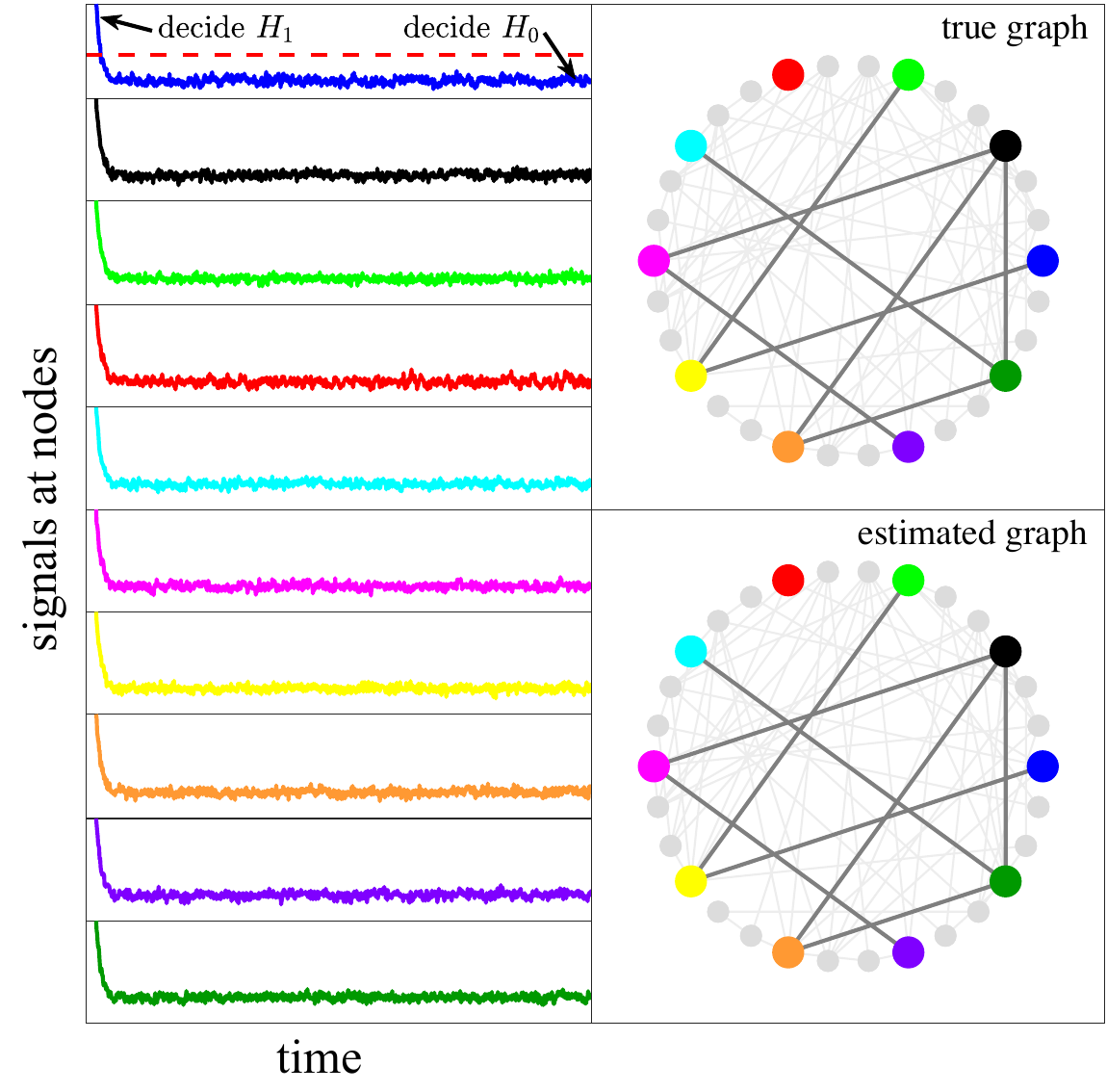}
\caption{Example of graph learning under distributed detection. 
The signals on the left represent the detection statistics evolving at the probed nodes, with color codes corresponding to the nodes highlighted in the topologies displayed on the right. 
The horizontal dashed line in the top-left panel represents a detection threshold set equal to zero. 
The dual learning problem (graph learning) is shown in the right panels, where the graph of probed nodes is correctly retrieved by the estimation algorithm.}
\label{fig:DAD}
\end{center}
\end{figure}

\begin{figure*}[t]
\centering
\[\begin{array}{cc}
\includegraphics[width=60mm]{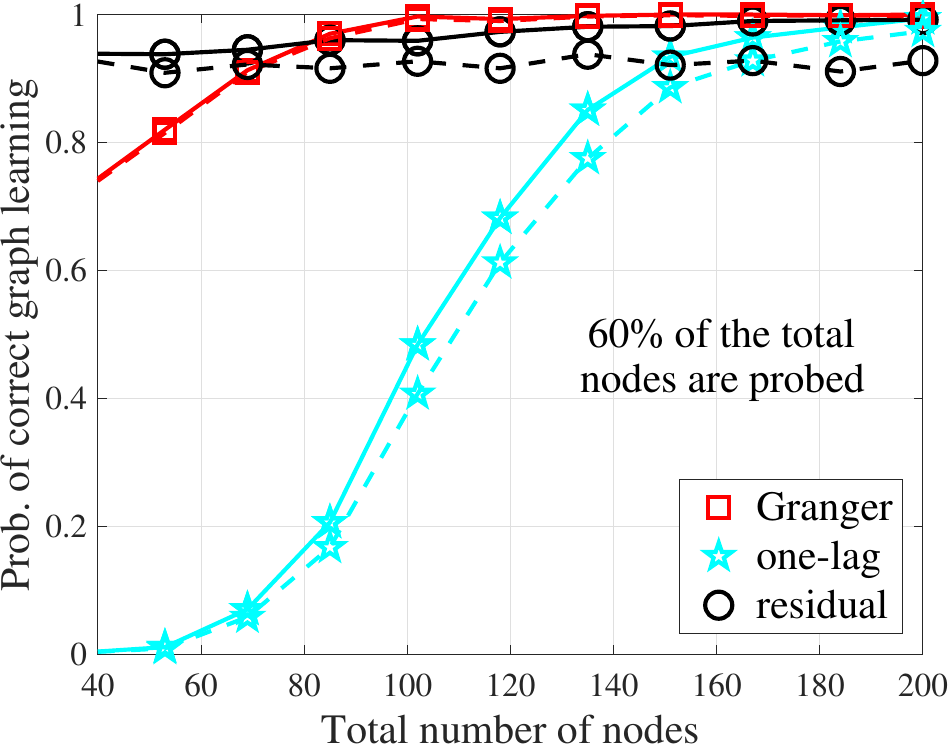}&
\includegraphics[width=60mm]{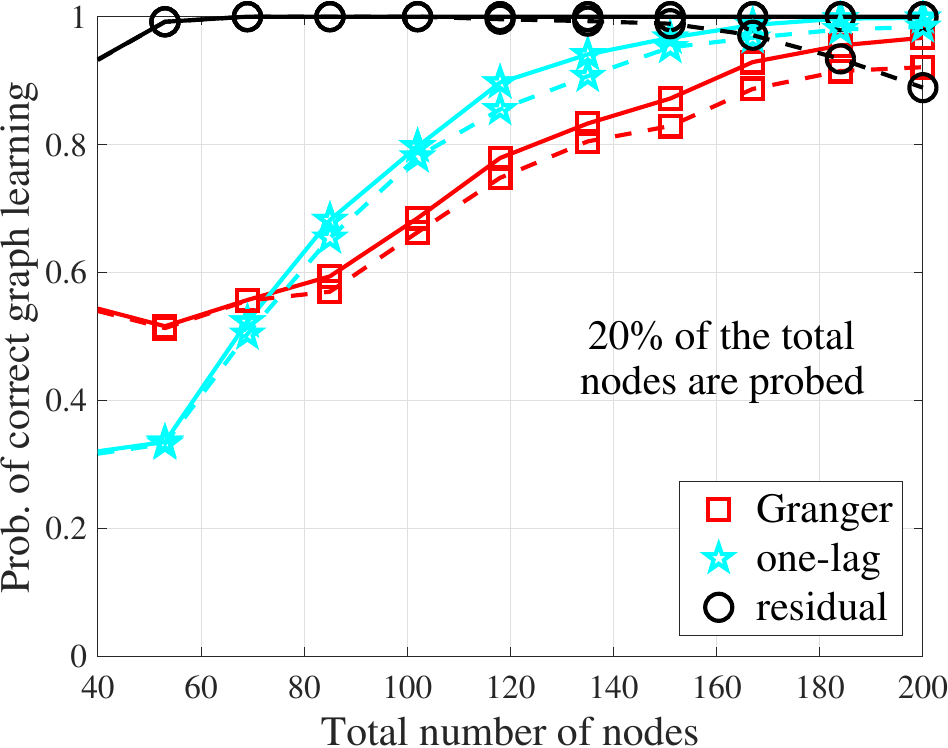}\\	
\includegraphics[width=60mm]{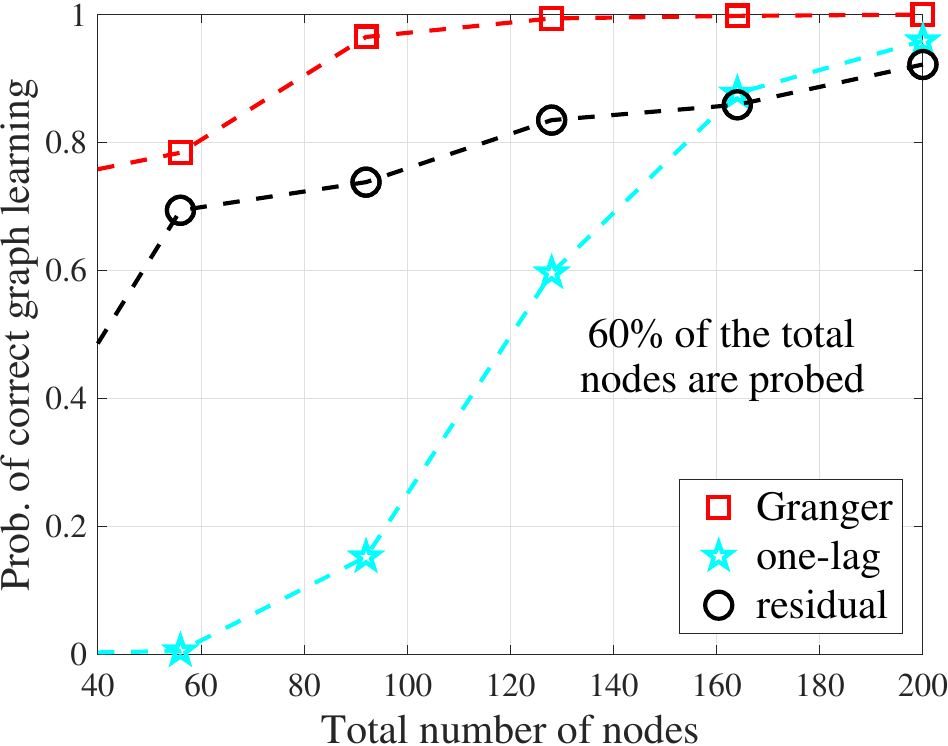}&
\includegraphics[width=60mm]{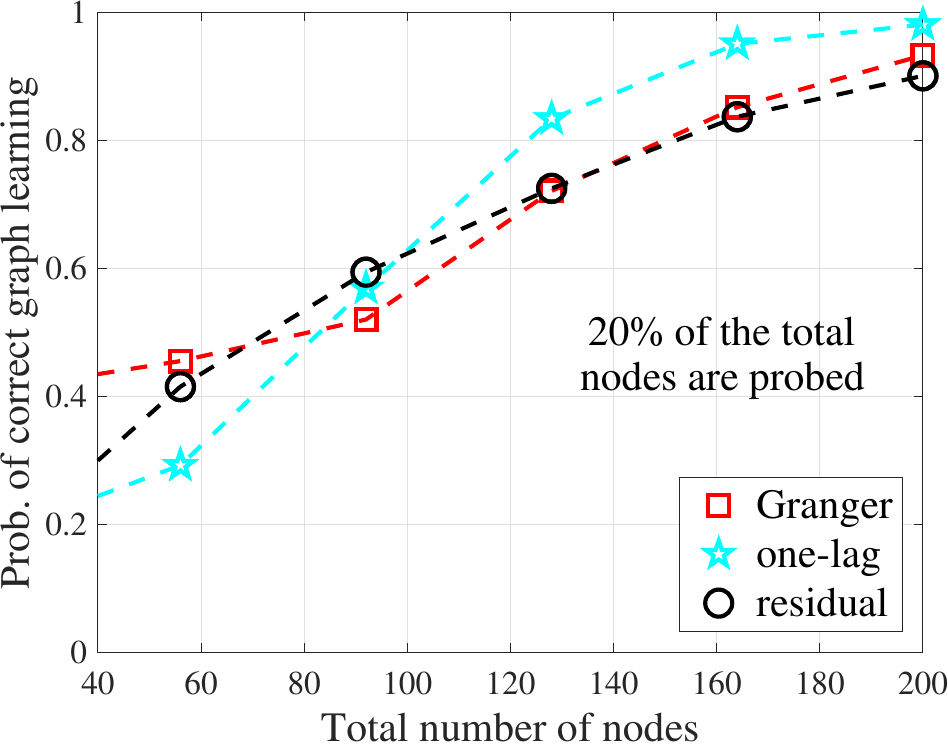}	
\end{array}
\]
\caption{Graph learning over the dynamical system in~(\ref{eq:mainmod}). 
The probability of correct graph learning is evaluated over $10^3$ Monte Carlo runs, where for each run the graph is generated according to a plain Erd\H{o}s-R\'enyi model with connection probability $p=0.1$. The input data are generated according to a standard Gaussian distribution, and the combination matrix has parameter $\rho=0.99$. 
Left panels refer to a Metropolis combination rule, right panels to a Laplacian rule with $\alpha=0.9$. 
In the top panels, solid lines refer to the limiting performance (i.e., unlimited samples), whereas dashed lines refer to a fixed number of $5\times 10^5$ samples. In the bottom panels, the number of samples is varied with the network size $N$ according to the scaling law $({\sf D}_{\textnormal{av}})^2\log S$, and is equal to $5\times 10^5$ in the last point ($N=200$). We recall that ${\sf D}_{\textnormal{av}}$ is the average degree of the network and $S$ is the number of probed nodes.
}
\label{fig:figsim}
\end{figure*}

In Fig.~\ref{fig:DAD}, we consider a network engaged in solving a Gaussian shift-in-mean detection problem, where the data are i.i.d. unit-variance Gaussian random variables whose mean is equal to $-1$ under the null hypothesis $H_0$, and to $+1$ under the alternative hypothesis $H_1$. 
The step-size of the CTA diffusion algorithm is set equal to $\mu=0.1$, and a Metropolis rule is employed to build the combination matrix.
We assume that all nodes initially (i.e., at time zero) believe that the true hypothesis is $H_1$, while, in contrast, the data that they start observing are actually generated according to $H_0$. 
In the network topologies on the two right panels, the nodes that can be probed are highlighted by different colors, whereas the nodes that are not accessible are displayed in gray. 
In the ten left panels, we display the output of the distributed detection problem (i.e., the direct learning problem), namely, the signals $\{\bm{y}_k(i)\}_{k\in\mathcal{S}}$ that are collected by the inferential engine in order to solve the graph learning problem (i.e., the dual learning problem). 
The color of the particular signal refers to the color of the corresponding node in the graph topology.

First, we see that the distributed detection algorithm is effective in accomplishing its task. 
In fact, after a relatively short transient all nodes' output signals fluctuate around the negative value $-\mathcal{D}_{01}$, which allows them to choose the correct hypothesis $H_0$ by using, e.g., a threshold equal to zero (horizontal dashed line in Fig.~\ref{fig:DAD}).

Second, despite the apparent similarity between the signals at different nodes, we see that there is significant information contained in these data streams about the node interactions, i.e., about the network subgraph in $\mathcal{S}$. 
As a matter of fact, graph learning is possible, as we can appreciate from the boxes on the right, which highlight the correct reconstruction of the subgraph of probed nodes (in particular, in this example the Granger estimator is used).

\begin{figure*} [t]
\begin{center}
\includegraphics[scale= 0.4]{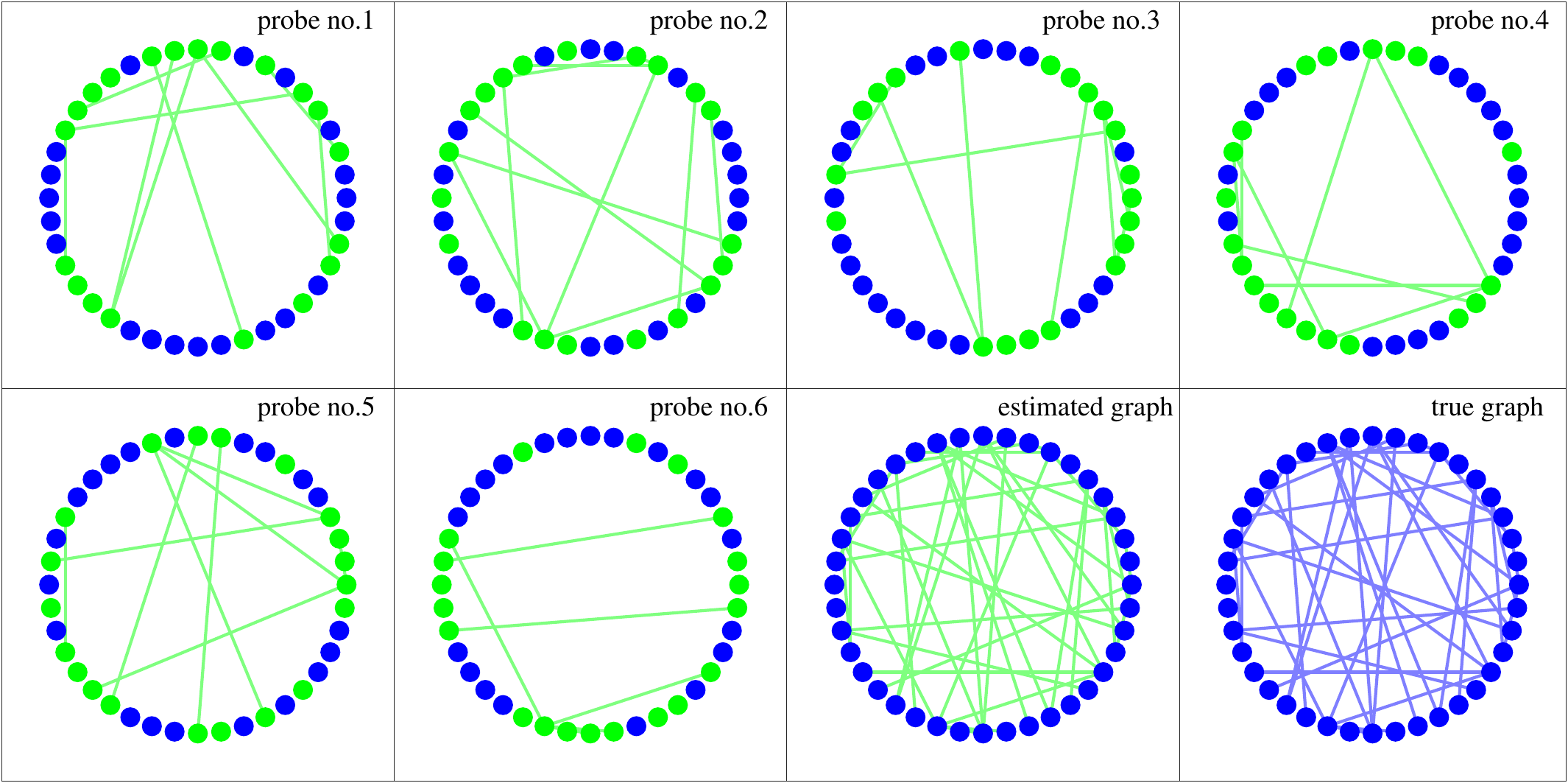}
\caption{Example of sequential graph learning over the dynamical system in~(\ref{eq:mainmod}), with standard Gaussian input data and Metropolis combination rule with parameter $\rho=0.5$. 
Successive {\em local} graph-learning experiments are shown, where each experiment corresponds to probing a subset of nodes (highlighted in green). 
For each probe, the {\em estimated} edges linking the subset of {\em currently probed nodes} are displayed in green as well. 
The matrix estimator used in the single experiments is the Granger estimator. 
In the last two (bottom) panels, we display (second to last panel) the {\em total} estimated graph learned by aggregating the six local experiments; and (last panel) the true graph. We see that the true graph is eventually learned.}
\label{fig:P&C}
\end{center}
\end{figure*}

\subsection{Performance, Complexity and Finite Sizes}
Let us examine the performance of the graph learning strategies. 
We consider the Granger, one-lag and residual estimators computed with both the exact covariance matrices and the empirical covariance matrices estimated from the samples.   
In Fig.~\ref{fig:figsim} we show the probability of exact recovery of the subgraph of probed nodes, for increasing network sizes and with fixed percentage of probed nodes. 
Two instances are considered, namely, one where the percentage of probed nodes is $60\%$, the other where the percentage is $20\%$. 
Consider first the top panels in Fig.~\ref{fig:figsim}. Here, the number of samples (which is relevant to the empirical estimators, dashed lines) is kept fixed across all values of $N$. 
Comparing the two top panels, some useful trends emerge. 
First, we see that in both cases, and for all the estimators, graph learning becomes effective as the network size increases, and this happens for moderate network sizes. 
Second, we see that the Granger estimator  is the best among the three estimators for $\xi=0.6$, but not for $\xi=0.2$. 
This behavior is interesting since it highlights that, even if the Granger estimator is exact in the case of full observability (which justifies why it works well for relatively high values of $\xi$), it can be outperformed under partial observability.  
Third, we see that in the top panels the residual estimator is almost the best one, even if it seems to be more sensitive to the number of samples (see dashed lines). We will get more elements on this behavior from the subsequent analysis.

The limiting performance with unlimited sample size (solid lines) is in principle attainable by the sample estimators (dashed lines) with sufficiently large number of samples.
However, as described in Sec.~\ref{sec:literature}, a relevant question is to determine how many samples are necessary. 
This aspect has been overlooked so far. 
A sample complexity analysis for the Granger, one-lag and residual estimators is performed in~\cite{tomo3}. 
The analysis reveals that the sample complexity is on the order of ${\sf D}_{\textnormal{av}}^2\log S$, which loosely ranges from quadratic in $N$ in the dense case, to less than linear in the sparse case.

According to this observation, in the bottom panels of Fig.~\ref{fig:figsim} we consider the same parameters of the top panels, but with a number of samples that grows with the network size, scaling as ${\sf D}_{\textnormal{av}}^2\log S$. 
Since in a sample complexity analysis we want to examine the impact that a reduced number of samples has on the learning performance, the number of samples used in the bottom panels is never greater than the number of samples used in the top panels. 
In particular, in the bottom panels the number of samples increases with $N$ and in the last point ($N=200$) is set equal to the number of samples used in the top panels.
Two notable effects emerge. First, we see that with this scaling law for the sample complexity, graph learning becomes effective as the network size grows. Moreover, while the performance of the Granger and one-lag estimators confirms the behavior (and relative ordering) seen in the top panels, the performance of the residual estimator does not, highlighting a major sensitivity to finite-samples effects.

\subsection{Sequential Graph Learning}
The results in Table~\ref{tab:summary} show that, under appropriate conditions, it is possible to estimate a subgraph by probing only the nodes in that subgraph, i.e., {\em locally}. This suggests that the entire network can be reconstructed through a {\em sequence of learning experiments that consider only small patches} of the overall network~\cite{tomo2,tomo_dsw}. 
This sequential scheme is of great interest over large networks, where one could eventually cover all nodes, but not simultaneously. For example, for various types of constraints (i.e., computation, accessibility) it might be impractical to measure all signals from the network. 
Nevertheless, by integrating the partial results coming from each patch, we can eventually estimate the entire graph. 

An example of this sequential reconstruction is offered in Fig.~\ref{fig:P&C}, where the boxes are numbered progressively to denote the current patches under test. 
For each probe, the graph learning algorithm produces an estimate of the subgraph (displayed in green) linking the currently probed nodes. 
In the shown example, we assume that the network is partitioned into a certain number of non-overlapping equal-sized patches. 
The overall ensemble of patches covers the whole network. 
Moreover, we consider that at each probe, a pair of these patches is chosen and that after all probes, all possible pairs are tested.
In the second to last bottom box, we display the {\em overall} network graph that is learned by aggregating the information relative to the individual patches. 
In the last bottom box we display the true graph. 
Comparing the latter two boxes, we see that the true graph is ultimately learned by the sequential reconstruction algorithm.

\section{Conclusions and Outlook}
\label{sec:conclu}
This article surveys state-of-the-art methods in the area of graph learning under partial observability. 
Under this setting, data from only a portion of the network is available, and the main question is: {\em Can the subgraph of probed nodes be properly estimated despite the presence of many latent unobserved nodes?} We described the challenges that arise in this context, and presented algorithms and performance limits that enable consistent learning under both sparse and dense graph regimes.

Several extensions are possible, such as considering higher-order and nonlinear dynamical models, directed graphs, asymmetric combination policies, and other random graph models~\cite{BarabasiAlbert}. Furthermore, examining sample complexity for more effective graph learning is an important question that deserves closer examination.


\end{document}